\documentclass[twocolumn]{aastex62}
\usepackage{amsmath}
\usepackage{url}
\usepackage{natbib}
\usepackage{hyperref}
\usepackage{lipsum}
\usepackage{xcolor}
\usepackage{multirow}
\usepackage{placeins}
\usepackage{braket}
\usepackage{outlines}
\usepackage{enumitem}

\setcitestyle{citesep={,}}

\newcommand{\PAPER}{\mathrm{PAPER}}
\newcommand{\hMpci}{h\ {\rm Mpc}^{-1}}
\shorttitle{Simplified analysis of PAPER-64}
\shortauthors{Kolopanis et al.}

\DeclareGraphicsExtensions{.eps, .pdf, .png}
\newcommand{\upperlims}{$(1500$ mK$)^{2}$, $(1900$ mK$)^{2}$, $(280$ mK$)^{2}$, $(200$ mK$)^{2}$, $(380$ mK$)^{2}$, $(300$ mK$)^{2}$ at redshifts $z=10.87,\ 9.93,\ 8.68,\ 8.37,\  8.13,$ and $7.48$, respectively}

\newcommand{\code}[1]{{\MakeUppercase{\footnotesize #1}}}

\defcitealias{ali_et_al2015}{A15}
\defcitealias{cheng_et_al2018}{C18}

\begin{document}

\begin{abstract}
We present limits on the  21cm power spectrum from the Epoch of Reionization (EoR)  using data from the 64 antenna configuration of the
Donald C. Backer Precision Array for Probing the Epoch of
Reionization (PAPER) analyzed through a power spectrum pipeline  independent from previous PAPER analyses.
 Previously reported results from PAPER
have been found to contain significant signal loss  \citep{cheng_et_al2018}.
Several lossy steps from previous PAPER pipelines have not been included in this analysis, namely:
 delay-based foreground filtering, optimal fringe-rate filtering, and empirical covariance-based estimators.
Steps which remain in common
with previous analyses include redundant calibration and local sidereal time (LST) binning.
The power spectra reported here are effectively the result of applying a
linear Fourier transform analysis to the calibrated, LST binned data.
This analysis also uses more data than previous publications, including the complete available
 redshift range of $z\sim7.5$ to $11$.
In previous PAPER analyses, many power spectrum measurements were found to be
detections of noncosmological power at levels of significance ranging from two to hundreds of times
the theoretical noise. Here,
%, with the expanded scope of more data and asignificantly reduced number of analysis steps, a more detailed study of thesedetections is possible.  
excess power is examined
using redundancy between baselines and power spectrum jackknives.
The upper limits we find on the 21cm power spectrum
from reionization are \upperlims.
For reasons described in \citep{cheng_et_al2018}, these limits supersede all previous PAPER results (\citealt{ali_et_al2015_erratum}).
\end{abstract}

\title{A simplified, lossless re-analysis of PAPER-64}

\author[0000-0002-2950-2974]{Matthew Kolopanis}
\altaffiliation{\href{mailto:matthew.kolopanis@asu.edu}{matthew.kolopanis@asu.edu}}
\affiliation{School of Earth and Space Exploration, Arizona State U., Tempe AZ}

\author{Daniel C. Jacobs}
\affiliation{School of Earth and Space Exploration, Arizona State U., Tempe AZ}

\author{Carina Cheng}
\affiliation{Astronomy Dept., U. California, Berkeley, CA}

\author{Aaron R. Parsons}
\affiliation{Astronomy Dept., U. California, Berkeley, CA}
\affiliation{Radio Astronomy Lab., U. California, Berkeley CA}

\author[0000-0001-6744-5328]{Saul A. Kohn}
\affiliation{Dept. of Physics and Astronomy, U. Penn., Philadelphia PA}

\author{Jonathan C. Pober}
\affiliation{Dept. of Physics, Brown University, Providence RI}

\author[0000-0002-4810-666X]{James E. Aguirre}
\affiliation{Dept. of Physics and Astronomy, U. Penn., Philadelphia PA}

\author{Zaki S. Ali}
\affiliation{Astronomy Dept., U. California, Berkeley, CA}

\author[0000-0002-0916-7443]{Gianni Bernardi}
\affiliation{INAF-Istituto di Radioastronomia, via Gobetti 101, 40129, Bologna, Italy}
\affiliation{Department of Physics and Electronics, Rhodes University, PO Box 94, Grahamstown, 6140, South Africa}
\affiliation{South African Radio Astronomy Observatory, Black River Park, 2 Fir Street, Observatory, Cape Town, 7925, South Africa}

\author{Richard F. Bradley}
\affiliation{Dept. of Electrical and Computer Engineering, U. Virginia, Charlottesville VA}
\affiliation{National Radio Astronomy Obs., Charlottesville VA}
\affiliation{Dept. of Astronomy, U. Virginia, Charlottesville VA}

\author{Chris L. Carilli}
\affiliation{National Radio Astronomy Observatory, Socorro, NM}
\affiliation{Cavendish Lab., Cambridge UK}

\author[0000-0003-3197-2294]{David R. DeBoer}
\affiliation{Radio Astronomy Lab., U. California, Berkeley CA}

\author{Matthew R. Dexter}
\affiliation{Radio Astronomy Lab., U. California, Berkeley CA}

\author[0000-0003-3336-9958]{Joshua S. Dillon}
\altaffiliation{NSF AAPF Fellow}
\affiliation{Astronomy Dept., U. California, Berkeley, CA}

\author{Joshua Kerrigan}
\affiliation{Dept. of Physics, Brown University, Providence RI}

\author{Pat Klima}
\affiliation{National Radio Astronomy Obs., Charlottesville VA}

\author[0000-0001-6876-0928]{Adrian Liu}
\affiliation{Department of Physics and McGill Space Institute, McGill University, Montreal, QC, Canada}
\affiliation{CIFAR Azrieli Global Scholar, Gravity \& the Extreme Universe Program, Canadian Institute for Advanced Research, 661 University Ave., Suite 505, Toronto, Ontario M5G 1M1, Canada}

\author{David H. E. MacMahon}
\affiliation{Radio Astronomy Lab., U. California, Berkeley CA}

\author{David F. Moore}
\affiliation{Dept. of Physics and Astronomy, U. Penn., Philadelphia PA}

\author[0000-0003-1602-7868]{Nithyanandan Thyagarajan}
\altaffiliation{Jansky fellow of the National Radio Astronomy Observatory}
\affiliation{National Radio Astronomy Observatory, Socorro, NM}

\author{Chuneeta D. Nunhokee}
\affiliation{Department of Physics and Electronics, Rhodes University, PO Box 94, Grahamstown, 6140, South Africa}

\author{William P. Walbrugh}
\affiliation{INAF-Istituto di Radioastronomia, via Gobetti 101, 40129, Bologna, Italy}

\author{Andre Walker}
\affiliation{INAF-Istituto di Radioastronomia, via Gobetti 101, 40129, Bologna, Italy}

\section{Introduction}
\setcounter{footnote}{0}
The Epoch of Reionization (EoR) represents a global phase
transition for intergalactic hydrogen from a neutral to ionized state.
In most models, this phase transition is fueled by the first
luminous bodies, which condensed from hydrogen clouds and
began heating and ionizing the surrounding Intergalactic Medium (IGM) \citep{barkana_loeb2001, Oh_2001}.
Observational constraints limit the timing of this event to somewhere
in the redshift range (12 $<$ z $<$ 6).

The 21cm photons emitted from the spin-flip transition of
hydrogen
are predicted to be a powerful probe of cosmic evolution
during this time \citep{furlanetto_et_al2006}. For in-depth
reviews of the physics of
21cm cosmology, refer to
\citet{barkana_loeb2007,morales_wyithe2010,loeb_furlanetto2013}
and \cite{pritchard_loeb2010}.

As observed from Earth, the 21cm line is redshifted
into the 100MHz radio band where it competes with
human interference and astrophysical emission from both the Milky Way and other galaxies. Interference is mitigated by
careful RF design and choosing a remote and regulated location for observation\footnote{PAPER was located at the South Africa Square Kilometer Array close to the current home of Meerkat.},
leaving astrophysical foregrounds as the principal contaminant,
 dominating the cosmological 21cm background by 4 or 5
orders of magnitude.
The foreground challenges faced by modern radio arrays are discussed
in detail in previous literature \citep[e.g.][]{santos_et_al2005,deoliveira2008,ali_et_al2008,
bernardi_et_al2009,bernardi_et_al2010,
bernardi:2013,ghosh_et_al2011,pober_et_al2013, yatawatta_et_al2013}.

Detection of 21cm emission by the neutral hydrogen medium is the target
of multiple experiments including those aimed at a globally averaged total power
measurement  (EDGES; \cite{bowman_et_al2010},
LEDA; \cite{bernardi_et_al2016}, %\cite{greenhill_bernardi2012},
SARAS; \cite{patra_et_al2015},
BIGHORNS; \cite{sokolowski_et_al2015}, and
SCI-HI; \cite{voytek_et_al2014})
and the fluctuations
caused by heating, cooling, collapse, and ionization
(GMRT; \cite{paciga_et_al2013},
LOFAR\footnote{\url{www.lofar.org}}; \cite{yatawatta_et_al2013},
MWA\footnote{\url{mwatelescope.org}}; \cite{tingay_et_al2013}, and
HERA\footnote{\url{reionization.org}}; \cite{DeBoer_et_al2016}).

The Donald C. Backer Precision Array for Probing the Epoch of Reionization (PAPER\footnote{\url{eor.berkeley.edu}}; \cite{parsons_et_al2010})
was an experimental interferometer with the goal of placing some of the first limits on these fluctuations.
The PAPER experiment observed in stages, with the number of antennas increasing by factors of two roughly every year.
Previous PAPER publications include
the 8 station results \citep{parsons_et_al2010},
 the 32 element power spectrum estimates (\citet{pober_et_al2013,parsons_et_al2014,jacobs_et_al2014}, \citet{moore_et_al2017}),
the 64 element power spectrum estimates (\citet{ali_et_al2015}; hereafter \citetalias{ali_et_al2015}),
 and our companion paper (\citet{cheng_et_al2018}, hereafter \citetalias{cheng_et_al2018}).

\begin{figure*}[tp]
\begin{center}
\includegraphics[width=\textwidth]{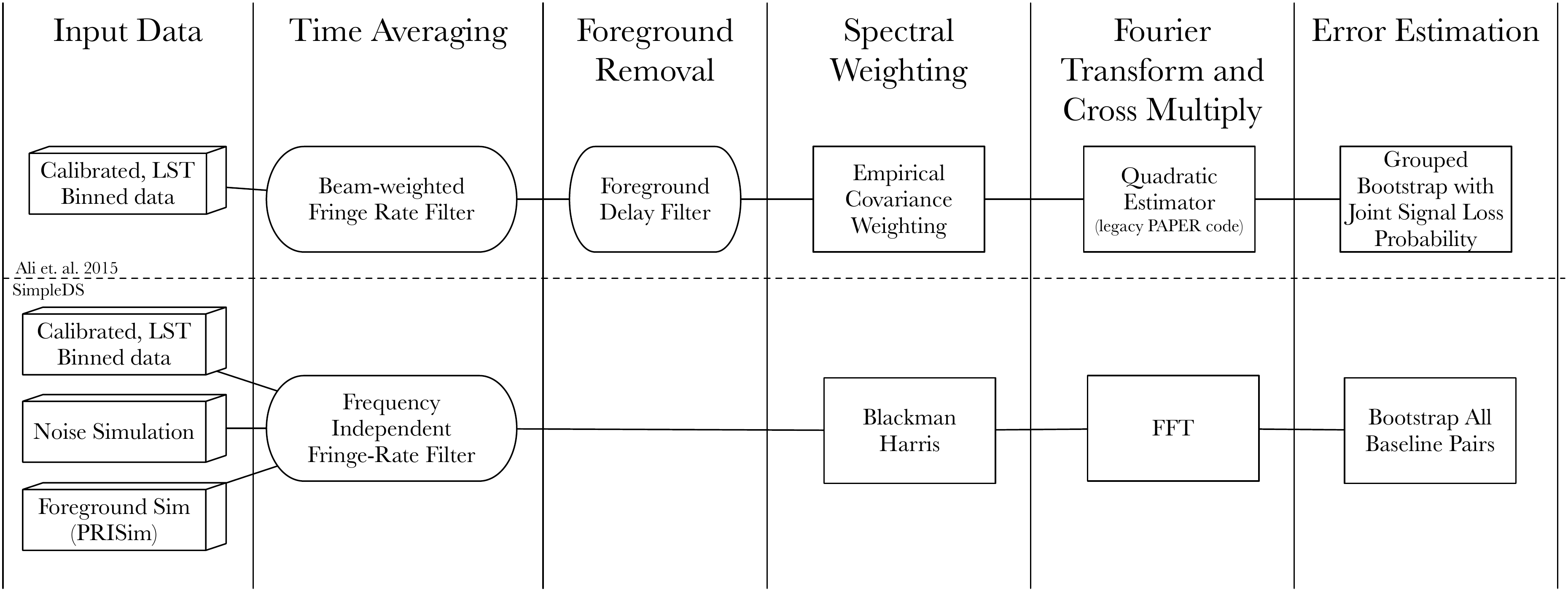}
\caption{Comparison between the prior PAPER analysis by \cite{ali_et_al2015} and ``simpleDS''.
The frequency independent fringe rate filter has a smoother
delay response compared to the one used in
\citetalias{ali_et_al2015} and
\citetalias{cheng_et_al2018} in order to reduce leakage
of foreground power outside the wedge.
The delay filter for foreground removal has been
omitted from this analysis to keep the pipeline
as simple as possible. While the foreground removal
technique should not affect cosmological
signals outside the wedge
\citep{parsons_backer2009, parsons_et_al2012b, parsons_et_al2014},
recent works have shown the use of this filter does
not produce a statistically significant reduction
in power at high delay modes \citep{kerrigan_et_al2018}.
Also, we find that the Fourier Transform from
frequency into delay is not dynamic range limited
when including the foreground signals.
Most importantly, in order to avoid signal loss during power spectrum
estimation, we use a uniformly weighted
Fast Fourier Tranform (FFT) estimator instead of the
empirical inverse covariance weighted OQE used in
previous PAPER works. }
\label{fig:pipeline_compare}
\end{center}
\end{figure*}

Through the re-analysis described in \citetalias{cheng_et_al2018}, additional signal loss in the empirical covariance
inversion method was discovered \citep{ali_et_al2015_erratum}.
Signal loss is the
unintentional removal of the target cosmological signal during analysis.
In \citetalias{ali_et_al2015}, this results from the use of empirically estimated covariance
matrices as a weighting matrix in the Quadratic Estimator (QE)
during power spectrum estimation.
An empirically estimated covariance matrix contains
terms related to the data, this dependence induces higher order (i.e. non-quadratic) terms in the estimator.
Applying QE normalization despite these terms then violates the assumptions of the statistics of the quadratic estimators and produces a biased results with incorrect power levels (e.g. signal loss).
This effect is described
more thoroughly in Section 3.1.1 of \citetalias{cheng_et_al2018}.
\citetalias{cheng_et_al2018} also describes how the amount of signal loss
in the \citetalias{ali_et_al2015} analysis was underestimated
and was further obfuscated by similarly underestimated uncertainties (from both analytic noise estimates and bootstrapped error bars). The \citetalias{cheng_et_al2018} analysis presents a detailed look
at the origin of these issues but does not deliver a revised analysis for the same data. In this paper we take a different look using an independently developed pipeline which conservatively has had many lossy steps removed (see Figure \ref{fig:pipeline_compare}).

Specifically, we aim to make improvements in two areas. First, we use the
independently developed pipeline \code{simpleDS}\footnote{\url{github.com/RadioAstronomySoftwareGroup/simpleDS}} which has minimal common code with the original PAPER pipeline built for \citetalias{ali_et_al2015} and
extended by \citetalias{cheng_et_al2018}.
 Second this analysis reduces the number of pipeline steps.  The basic concept of the delay spectrum
is retained with a power spectrum measurement coming from each type of baseline; however several steps have been removed and others replaced.
The steps used in this type of analysis can be broken into three sections: calibration and averaging over multiple nights (LST binning), foreground filtering and time
averaging, and power spectrum estimation.

The re-analysis described in \citetalias{cheng_et_al2018} focused almost exclusively on
the final stage. In this analysis the intermediate stages (like foreground filtering) have been re-examined, and in
all cases either removed or simplified.
This paper uses data sets which have
been previously interference flagged, calibrated with redundant calibration,
LST binned, and absolutely calibrated to Pictor A.
As this analysis takes advantage of archival LST binned data products,
the stages prior to binning are unchanged from previous analyses.

This paper is organized as follows: we discuss the three pipeline
inputs by reviewing the data used
in this analysis in Section~\ref{sec:data},
the input noise simulation in Section~\ref{sec:noise_sim}, and the simulated sky input used to calibrate power spectrum normalization and examine
additional signal loss in Section~\ref{sec:sims}. The major changes in the analysis pipelines
between this work an \citetalias{ali_et_al2015} are discussed in Section~\ref{sec:pipeline_comparison}.
We investigate how closely the PAPER baselines adhere to the redundant layout in Section~\ref{sec:redundancy}.
In Section~\ref{sec:OQE} we review
the revised power spectrum estimation techniques and uncertainties.
The multi-redshift power spectrum results are
presented in Section~\ref{sec:pspec_results} and upper limits on the 21~cm power spectrum are presented in Section~\ref{sec:upperlims}. Finally, we provide
some concluding remarks in Section~\ref{sec:conclusion}.

\begin{figure*}[tp]
\centering
\includegraphics[trim={3cm 0  4cm 0},width=\textwidth]{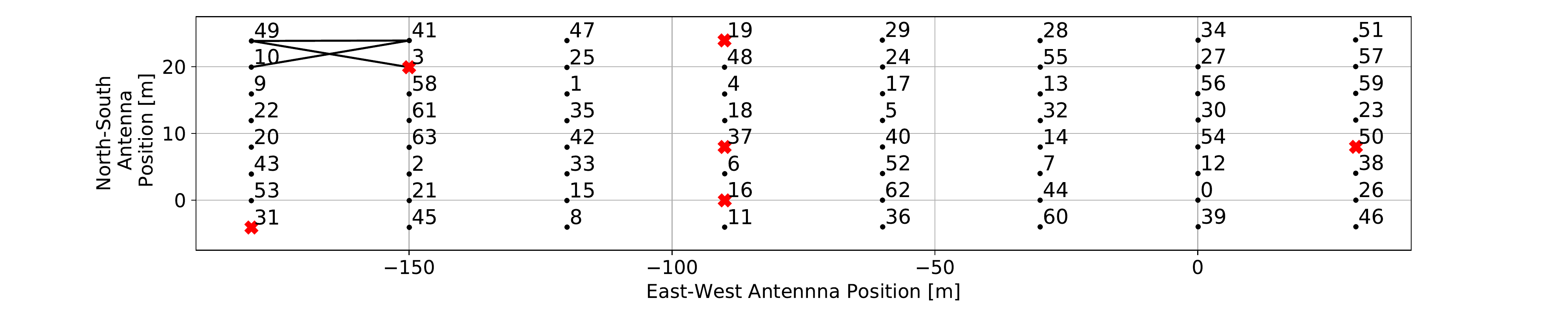}
\caption{The antenna positions of PAPER-64.
Highlighted are the three baseline types used in this analysis.
These baselines consist of East-West baselines from adjacent
antenna columns with no row separation
(e.g. 49-41, 1-4, 0-26),
baselines with one column separation and one positive Northward
row separation (e.g. 10-41, 1-48, 0-38),
and baselines with one column separation
and one negative Northward row separation
(e.g. 49-3, 1-18, 0-46). A red `x' denotes antennas which have been flagged from analysis. Reasons for
flagging include previously known spectral instability (19, 37, and 50), low number of counts in LST binning (3 and 16), and suspected non-redundant information (21 and 31).} \label{fig:ant_pos}
\end{figure*}

\begin{figure*}[t]
\centering
\includegraphics[trim={4cm 0  5cm 0},width=\textwidth]{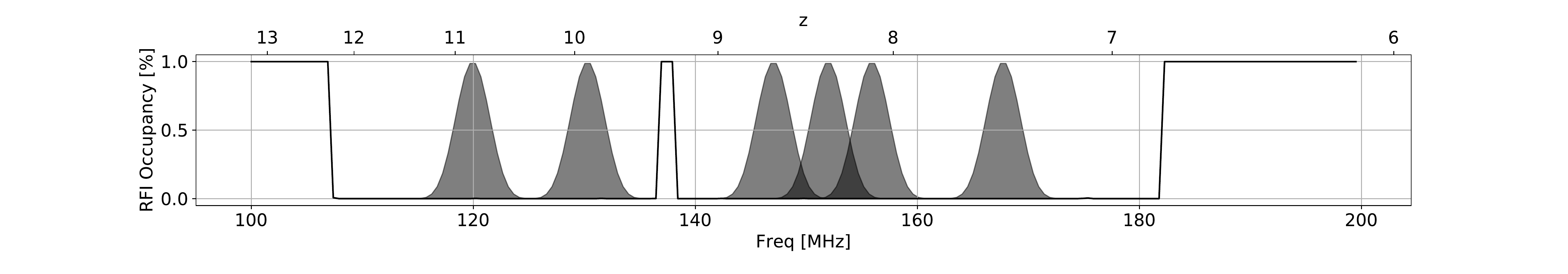}
\caption{The six frequency bands used in this analysis plotted over the
relative occupancy of flags from RFI.
Redshift bands are denoted by the
Blackman-Harris window functions used during
the Fourier transform from frequency to delay in order
to reduce foreground leakage to high delays.
The specific windows chosen here are centered on
$z=10.87,$ $9.93,$ $8.68$, $8.13,$ and $7.48$ ($119.7$, $130.0$, $146.7$, $155.6$, and $167.5$~MHz respectively).
Two sub-bands centered at $ 112 $ and $ 178 $~MHz could also be constructed with minimal RFI flagging; however, these bands contain significant high delay systematics even after the application of the Fringe-Rate Filter (FRF) and provide little unique information both cosmologically and towards the identification of persistent systematics. The model of the beam is dominated by extrapolation in some or all the frequencies in theses sub-bands and as a result and data products which depend heavily on the beam (the input simulation, thermal noise estimate, and input noise simulation) are not credible outside of the selected bands.
Frequency bands used in this analysis include the  150~MHz, $z=8.37$, band used in \citetalias{cheng_et_al2018} and \citetalias{ali_et_al2015}.
This redshift bin is included in order
to properly compare with previous works,
but it is worth noting the information
obtained from this bin is not entirely independent from the two
redshift bins with which it overlaps.
\label{fig:freq_select}}
\end{figure*}

\section{Data}\label{sec:data}
In the next three sections, we discuss the three major inputs to
our power spectrum pipeline: the observed data, simulated thermal noise, and the simulated foreground visibilities in Sections~\ref{sec:data},~\ref{sec:noise_sim},~and~\ref{sec:sims}, respectively.
\subsection{Data Selection}
\label{sec:data_selection}
The PAPER-64 antennas were arranged in an
 $8\times8$ grid as illustrated in Figure~\ref{fig:ant_pos}.   The grid arrangement
enables many repeated measurements of a single spatial Fourier mode
 to be averaged together before squaring, which delivers
higher sensitivity for these PAPER elements than a 
non-redundant configuration \citep{parsons_et_al2012a}.
This configuration is also well matched to the
delay spectrum method of measuring the power spectrum where visibilities are
Fourier transformed along the spectral dimension to make a one dimensional
 slice through the three dimensional Fourier domain \citep{parsons_et_al2012b}.

In principle, the delay spectrum method can be used to approximate a power spectrum for every pair of antennas,
which allows a great deal of freedom to explore systematic effects that vary from antenna pair to antenna pair. However, in this analysis, we limit our data volume by only forming power spectra from select baselines.
Specifically, we use only three baseline types of the shortest length ($30$~m)
as illustrated in Figure~\ref{fig:ant_pos}. The
shortest baselines are the most numerous and therefore provide the most sensitive measurements.
The shortest baselines also probe what are likely to be the brightest modes of the
diffuse reionization power spectrum.  However, the shortest spacings are also sensitive to diffuse foreground power which
is known to be brighter than the extragalactic point source background on these scales \citep{beardsley_et_al2016}.
 The exact tradeoffs between foregrounds, calibration error, and sensitivity are
 a matter of ongoing research.

The data used here comes from the PAPER-64 season which ran
for 135 nights between 2012 November 8
(JD 2456240) and 2013 March 23 (JD 24563745).
 Three antennas (19, 37, and 50) have been flagged due to higher levels of spectral instability and were also
flagged in \citetalias{ali_et_al2015}.

\subsection{Calibration and LST Binning}\label{sec:data_calibration}
The analysis described here begins with data that were previously compressed, calibrated, and LST binned.
 The details of the compression, calibration, and binning process are described more completely in \citetalias{ali_et_al2015}; here we briefly describe the salient details.
 Compression is achieved with the application of a Fringe-Rate Filter (FRF) (described in more detail in Section~\ref{sec:frf}) and a wide-band iterative deconvolution algorithm (WIDA) (described in more detail in Section~\ref{sec:wida}) to limit the data to fringe rates less than $ f \lesssim 23 $~mHz and delays less than $ |\tau| \lesssim 1 $~$ \mu$s. It also decimates along both time and frequency axes
 to Nyquist sample the data from the correlator output. These values are the same as \citetalias{ali_et_al2015} and the 
 compression process is described in more detail in \citet{parsons_et_al2014}.
 This compression process may imprint systematic biases in the data but are not investigated in this work.
After compression, data were first calibrated redundantly
using logarithmic calibration and linear calibration
techniques \citep{liu_et_al2010,zheng_et_al2014,dillon_et_al2017}.
An imaging-based flux density calibration was also
 applied using Pictor A fluxes derived from \citet{jacobs_et_al2013}.

The data are then grouped into bins according to local sidereal time.  Within each bin samples
with modified z-scores above $\sim4.5$ are flagged.
As opposed to z-scores, which use a sample set's
mean and standard deviation to find outliers,
modified z-scores use the median and
Median Absolute Deviation (MAD).
Modified z-scores are discussed in more
detail in Section~\ref{sec:redundancy} and thoroughly
in \citet{Iglewicz_and_hoaglin}.
 Data are binned into two sets one containing
odd numbered days and the other even.  These can then be differenced to estimate noise and cross
 multiplied for a power spectrum unbiased by noise.

\subsection{Flagging and Sub-band Selection}
We find that compared to all other antennas in the LST binned data set,
antennas 3 and 16 have an anomalously low number of samples.
After LST binning, most baselines have samples from between 30 and the full 64 days
in each frequency/time bin; baselines
associated with antennas 3 and 16 contain bins with
as few as 10 days sampled during the transit of Fornax A ($ \sim 3$~hours in LST).
 In the
interest of uniformity, these two antennas were therefore flagged and
excluded from analysis.
In a similar way, we limit the range of LSTs
included in the final power spectrum to times which are sampled repeatedly
throughout the observing season, corresponding to a time window between LSTs
 $00^{h}30^{m}00^{s} - 08^{h}36^{m}00^{s}$.\footnote{Note that the LST range here is slightly different from \protect\citetalias{ali_et_al2015} but is identical to the one used in \citetalias{cheng_et_al2018}. }

The data are then divided along the frequency axis into smaller redshift bins for further power spectrum analysis.
A practical limitation in redshift selection
comes from a desire to avoid including channels with significant RFI flagging.
Bands with the most continuous spectral sampling span the redshift range $11$ to $7.5$.
We select redshift ranges that are approximately coeval, i.e., bandwidths over which limited evolution of the 21 cm signal is expected.
%Constraints from EDGES high suggest that the evolution is probably slower than d$z$ (total change in
%redshift) of 1-2 \citep{monsalve_et_al2017} which corresponds to a spectral width of $ 26 $~MHz at 200~MHz and
%12 at 100~MHz. 
 To accommodate this constraint, adopt a band size of $10$~MHz.

This band size allows us to choose a number of spectral windows with very little to no RFI flagging.  The specific windows chosen here are centered on
$z=10.87,$ $9.93,$ $8.68$, $8.13,$ and $7.48$ ($119.7$, $130.0$, $146.7$, $155.6$, and $167.5$~MHz respectively).
These bands are illustrated visually in Figure~\ref{fig:freq_select}.
Two sub-bands centered at $ 112 $ and $ 178 $~MHz could also be constructed with minimal RFI flagging; however, these bands contain significant high delay systematics even after the application of the Fringe-Rate Filter (FRF) and provide little unique information both cosmologically and towards the identification of persistent systematics. The model of the beam is dominated by extrapolation in some or all the frequencies in theses sub-bands and as a result and data products which depend heavily on the beam (the input simulation, thermal noise estimate, and input noise simulation) are not credible outside of the selected bands.
As a validation check, we also include
a reprocessing of the $z=8.37$ bin centered at $151.7~$MHz
which was analyzed in \citetalias{ali_et_al2015} and \citetalias{cheng_et_al2018}.

\section{Noise Simulation}\label{sec:noise_sim}

In parallel with the observed PAPER data, we process a simulation of thermal noise
to help validate the simpleDS
pipeline's normalization, power
spectrum estimation, and bootstrapped variance
estimation techniques.
To generate the input noise simulation,
we assume the per baseline noise is
drawn from a complex Gaussian distribution $ \mathcal{CN}(0,\sigma_{n}) $.
To determine the width, $ \sigma_{n} $,
of this distribution,
we use the radiometer equation (9-15)
from \citet{clark1999}
\begin{align}
\sigma_{n}^{2} &= \frac{SEFD^{2}}{2\eta^{2} \Delta \nu t_{acc}} \label{eqn:radiometry}
\end{align}
where $ SEFD $ is the system equivalent
flux density, $ \eta $ is the antenna
efficiency, $ \Delta\nu $ is the observing
bandwidth in a frequency bin, and $ t_{acc} $ is the accumulation time of the observation.

The quantity $ \frac{SEFD}{\eta}$
is a measure of the expected variance of samples
of the total noise power.
Assuming the noise is Gaussian, the noise power
is the variance of the underlying distribution,
often described by a system temperature, $ T_{sys} $.
This quantity then is a measure of the
variance of the sample variance of a Gaussian
distribution, which equates to
$ \sigma_{n}^{2} \propto 2T_{sys}^{2} $. This factor of $ 2 $ will
 cancels with the factor in Equation~\ref{eqn:radiometry}.

Substituting this into Equation~\ref{eqn:radiometry}
yields an expression for the variance of a
realization of noise
\begin{align}
  \label{eqn:radiometry_tsys}
\sigma_{n}^{2} &= \frac{T_{sys}^{2}}{\Delta \nu t_{acc}  N_{days}}
\end{align}
where we have added the term $ N_{days} $ to account for the
averaging of individual samples during LST binning,
assuming the noise is independent between days.

We assume the system temperature, $ T_{sys} $ can be described by
the relations from \citet{rogers_bowman2008}
\begin{align}
T_{sys} = 180~\text{K}\left(\frac{\nu}{180~\text{MHz}}\right)^{-2.55} + T_{rcvr}
\end{align}
where we retain the parameters as measured or noted in past PAPER reports, most recently by \citetalias{ali_et_al2015}: a sky temperature model of $ T_{180}=180 $~K with a spectral index of $ \alpha=-2.55 $, a frequency independent receiver temperature, $ T_{rcvr} =144$~K (this parameter is taken from \citetalias{cheng_et_al2018}), a resolution of  $ \Delta\nu=\frac{100~\text{MHz}}{203} $ and an integration time $ t_{acc}=42.95 $~s.

Using Equation~\ref{eqn:radiometry_tsys}, we create a data set of Gaussian random noise matched in shape to the observed PAPER data.
This simulated noise data is processed through simpleDS in parallel with the PAPER data.

\section{Simulated Sky}\label{sec:sims}
There are several challenges in
making an accurate simulation of 21cm instruments, ranging from limited accuracy of catalogs to
the computational challenges in simulating large fields of view and large bandwidths. A simulation of millions of
sources from horizon to horizon over hundreds to thousands of channels and baselines is a formidable
challenge. Simulators addressing these challenges include
PRISim\footnote{The Precision Radio Interferometry Simulator (PRISim) is publicly available at \url{github.com/nithyanandan/PRISim}} \citep{prisim},
OSKAR\footnote{\url{https://github.com/OxfordSKA/OSKAR}} \citep{mort2010oskar},
FHD\footnote{\url{https://github.com/EoRImaging/FHD}} \citep{sullivan_et_al2012},
 and to a limited extent CASA\footnote{\url{https://casa.nrao.edu/}} \citep{casa_2007}. The pyuvsim\footnote{\url{https://github.com/RadioAstronomySoftwareGroup/pyuvsim}} python package
is currently being developed to produce exactly such simulations as well.

Testing the power spectrum code on a foreground-only instrumental simulation can reveal internal
inconsistencies, including scaling errors and other code errors; it can also help provide estimates of uncertainties in calibration and other sources of error.  $\PAPER$'s
wide field of view ($\sim 45^{\circ}$ full width half max beam with significant sensitivity all the way to the horizon \citep{pober_et_al2012}) drives
a requirement for a simulation which does not employ flat sky assumptions or approximations. One such simulator
is PRISim, which performs a full-sky visibility calculation given lists of catalogs \citep{nithya_et_al2015a,nithya_et_al2015b}. Using PRISim, we generate $ \sim8 $ hours of simulated $ \PAPER $ data matching the observing parameters of the LST binned data set.

The goal of this simulation is not to produce an accurate model of the sky suitable for subtraction or calibration,
but rather to provide a sky-like input to the
simpleDS power spectrum pipeline for power spectrum internal checks and rough comparison.
The simulation can confirm that overall scale of our final power spectrum
and helps identify
sky-like modes which may leak outside of the horizon under a
the delay transformation.

The sky model used by PRISim
 includes a GSM diffuse model (\citealt{deoliveira2008}),
point sources from the GLEAM sky survey with flux density $>1$~Jy at $150$ MHz
\citep{wayth:2015,hurley_walker2017}, a model of
Pictor A created from GLEAM, and a model of
 Fornax A created by
 using clean components derived from the deconvolution techniques described in \cite{sullivan_et_al2012} (Personal communication, Carroll \\\
  Byrne 2018).
 This Fornax A model has a total flux of $ 541.7 $~Jy at $ 180 $~MHz, consistent with the low frequency observations
 assuming a spectral index of $ -0.8 $ \citep{mckinely_et_al2015}.
PRISim simulates diffuse emission as collections of Gaussian point sources, much like
CLEAN components. The GSM component list is generated by interpolating the GSM HEALPix
map to be oversampled by a factor of 4 and each pixel
is then treated as an independent point source.

It is expected that this simulation will not perfectly reproduce the PAPER data,
due not only to incompleteness in the sky model and imperfections in the instrument model, but also because of potential errors or approximations in the methodology simulation code itself
(e.g. the choice to model the sky as made of point sources).
To avoid over-interpretation of the simulation results, we
limit our use of PRISim foreground simulations to
checking the flux scale (Section~\ref{sec:prisim_scale}),
understanding the impact of time averaging (Section~\ref{sec:common_mode}),
computing foreground error bar components (Section~\ref{sec:FG_err}),
constraining the general shape of the foreground power spectrum (Section~\ref{sec:pspec_results}),
and establishing the expected change in foreground power with LST (Section~\ref{sec:nulls}).

\subsection{Simulation Results}
We begin the comparison of the input data and simulated PRISim data by
noting that PRISim has, in the past, been primarily used to simulate delay power spectra rather than in the image domain; as such, we omit any detailed comparison of simulated phases with data.
Similarly, the PRISim implementation of the PAPER beam has not been tested at a detailed level (for example by imaging), and so delay modes near the horizon limit are not expected to be simulated as accurately as those well within the foreground wedge (see e.g. \citealt{pober_et_al2016})

\begin{figure}[tp]
\centering
\includegraphics[width=.5\textwidth]{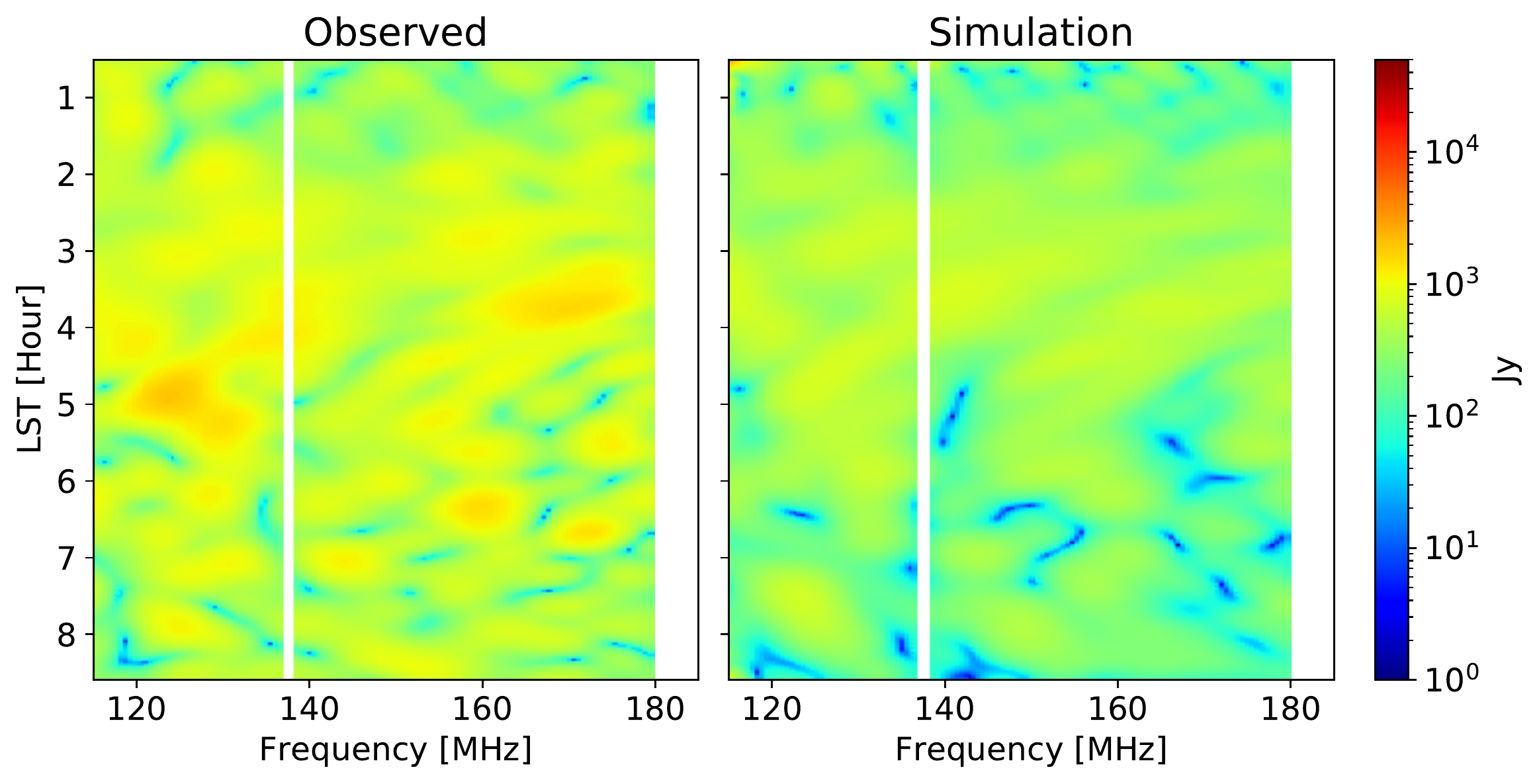}
\caption{An LST, frequency plot of the amplitude of a representative observed visibilities (left) and
the PRISim simulation (right) for the $ \sim 8$ hours
of data analyzed in this work.
While many details in the visibility amplitude structure do not match, the
there is general agreement, particularly near LST $ \sim3 $ hours when
Fornax A transits over the instrument.}
\label{fig:data_sim_waterfall}
\end{figure}

A comparison of the simulated and observed data is shown in Figure~\ref{fig:data_sim_waterfall}. Though some of the detailed fringing
structures are not reproduced in the simulation, 
the relative shape of the fluctuations appear
well-matched between the two data sets. The overall amplitude, however, 
of the two data products differs significantly.

\subsection{Absolute Calibration Check}\label{sec:prisim_scale}
One key question is the absolute calibration of the power spectrum amplitude scale. This scale
combines a number of factors including the absolute calibration performed to the data described in Section~\ref{sec:data_calibration}, the conversion from Jansky to mK, the Fourier transform convention, and
the cosmological scaling of delay modes.  Each is relatively simple but important to check (for example an error in $h$ scales as $h^3$ on $\Delta^2$).

\begin{figure*}[tp]
\centering
\includegraphics[width=.45\textwidth]{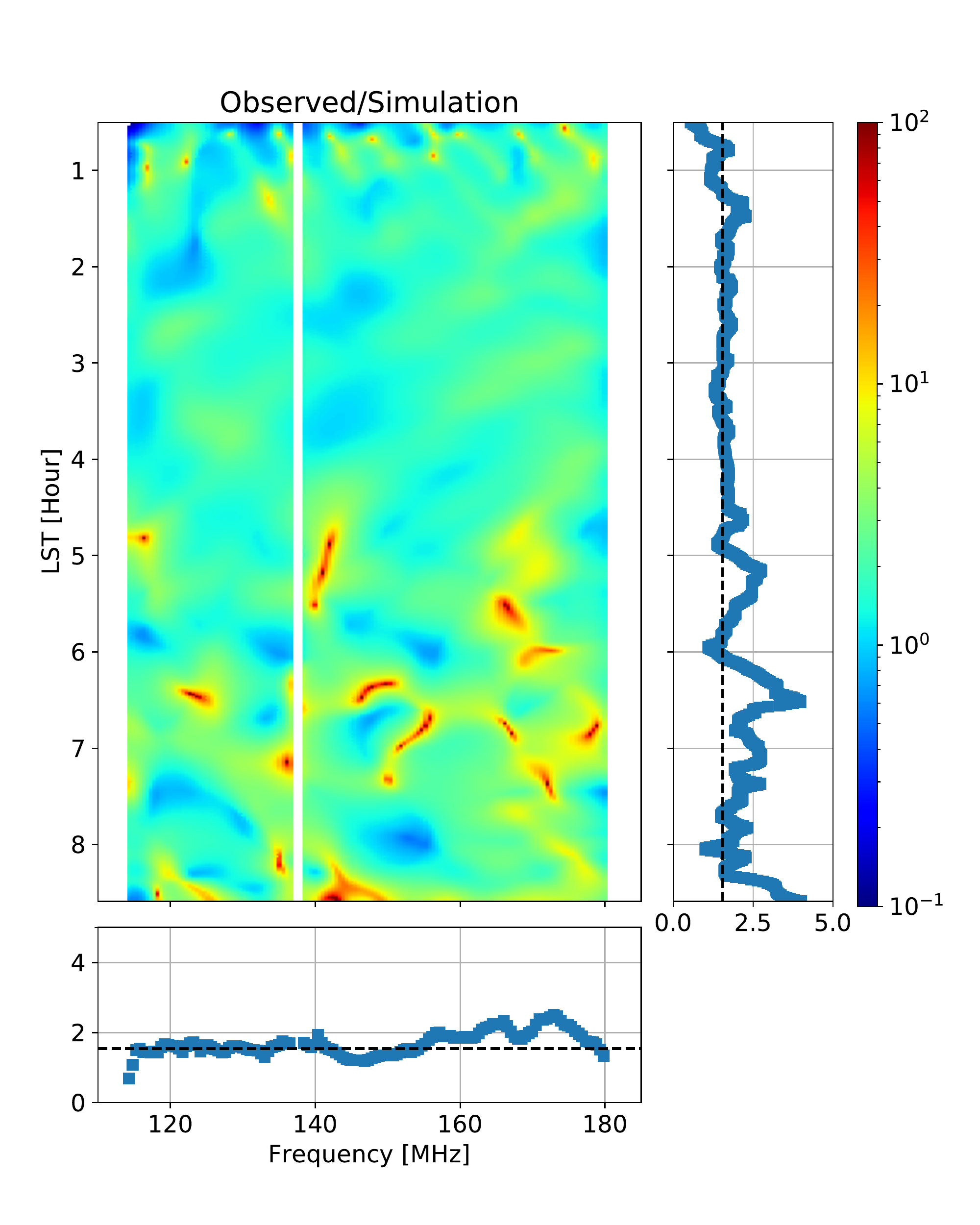}
\includegraphics[width=.45\textwidth]{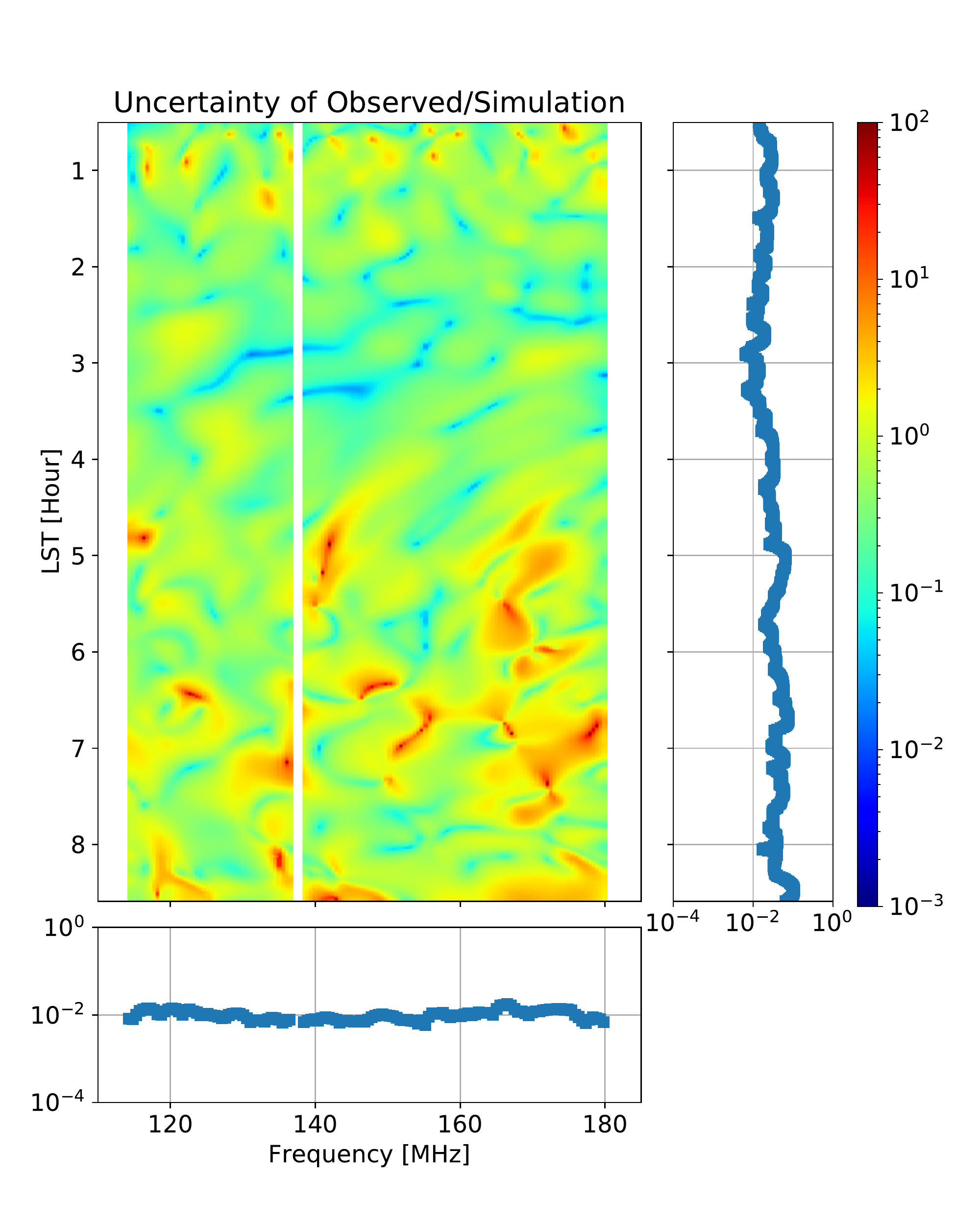}
\caption{The ratio of observed (left) and uncertainty (right) to  simulated visibility amplitude notice the difference in color scales. The observed is obtained by an unweighted averaged
over all baselines and the uncertainty from the variance across baselines.
Also plotted are the mean of the ratio and uncertainty %and standard deviations
averaged along the time axis (bottom panel)
and frequency axis (right panel) with the
maximum likelihood scale factor overplotted (black dashed).
While both the observed and simulated
data exhibit similar fringe patterns, the largest
differences occur between LSTs of 5 and 7 hours. This
is near the galactic anti-center and may be
indicative of an incomplete sky model. The most likely model scale factor
is $ 1.54 \pm 0.04 $ at $ 95\% $ confidence.
}
\label{fig:data_sim_ratio}
\end{figure*}

Figure~\ref{fig:data_sim_ratio} shows the ratio of amplitudes between observed and simulated visibilities, averaged over redundant baselines.  While both the observed and simulated
data exhibit similar fringe patterns, the largest
differences occur between LSTs of $ 5 $ and $ 7 $ hours.
This is near the galactic anti-center and may be
indicative of an incomplete sky model.
The PAPER beam model used is a polynomial fit of the spherical harmonic coefficients ($ a_{lm} $) fit from lab measurements taken between $ 120$ and $180 $~MHz; beyond
this range the simulated data is excluded from further analysis.

The ratio also becomes large where interference between fringing sources
drive the visibility amplitudes close to zero, but overall the ratio is generally close to unity.
These zero crossings make
a mean value difficult to interpret;
here we make a best estimate by computing the likelihood of a range of scale values ($g$) given the
baseline to baseline variation.

\begin{equation}
\log \mathcal{L}(g) = \sum_{\nu,t,bl} \frac{-(g * V(\nu,t)_{sim} - V(\nu,t)_{bl})^2}{2 * var(V(\nu,t)_{bl})}
\end{equation}
where the subscript $ bl $ refers to a unique redundant group, and the variance ($ var(V(\nu,t)_{bl} $) is computed over all baselines in a redundant group.

The scale factor is fit over the domains $ [.5, 4.5] $~hours in LST 
when the foreground simulation fringe pattern shows the most agreement with the observed visiblities and over the frequencies $ [120, 180] $~MHz where the PAPER beam model is most reliable.

The maximum likelihood scale factor is $1.54 \pm 0.04 $ at $ 95\% $ confidence. This is consistent with
the ratio observed during the first half of the data set in LSTs 1h to 5h (the dashed line plotted in Figure~\ref{fig:data_sim_ratio}).  This scaling factor
is used when estimating the expected foreground signal
in Section~\ref{sec:FG_err} and as an overall scaling
factor on the power spectrum estimated from the simulated data in Section~\ref{sec:pspec_results}.

\subsubsection{Model Scale Discussion}
The 50\% difference in scale between the model and the data is notable enough to merit
further discussion.  There are many possible sources for this difference, including uncertainty in catalog inputs to the PRISim simulator, the instrument model itself, the calibration of the
PAPER data, or some combination of all three. Deeper investigation requires careful testing of each component
separately, work which is beyond the scope of the present study. However, it is worth reviewing some of these aspects.

The original absolute calibration reported in \citetalias{ali_et_al2015} was done by imaging the Pictor field (at LST $=4$h) in each channel, correcting for a primary beam model and fitting a Gaussian to the extracted Pictor A source. This was done in ten minute snapshots with the resulting spectra averaged together. The standard deviation of the flux estimate was of order $\sim25$~Jy at $ 68\% $ confidence on each channel. A similar scale variation seen from channel to channel was consistent with sidelobe confusion. The change in scale due to that effect was on the order of a few percent.

The difference could also be attributed to calibration of the simulator.
Work is in progress to better verify the accuracy of array simulation codes;
lacking firm conclusions, we only expect PRISim simulations of diffuse structure to be accurate in amplitude to within a factor of two \citep{nithya_et_al2015a}.

Since the flux calibration of the simulations has not been rigorously independently tested, and the flux scale for the data is tied to a well-established model in \citet{jacobs_et_al2014}, we have scaled the simulation to match the data.  The flux calibration in this paper is thus unchanged from \citetalias{ali_et_al2015}.

%Given the uncertainties with regard to these possible sources of discrepancies,
%the overall scale difference observed between the
%simulated and observed data is perhaps not surprising.
%
%Since the overall amplitude of the observed data is on average greater than the simulation, we apply the scale 
%factor to increase the flux in the simulated data. We do not, however, use this scale factor to alter the 
%power spectrum estimates from the data.
%The flux calibration in this paper is unchanged from \citetalias{ali_et_al2015}

\section{Analysis Pipeline Comparison}\label{sec:pipeline_comparison}
In this section, we describe the differences in the analysis steps prior to Fourier Transform and power spectrum estimation
 between this work and \citetalias{ali_et_al2015}: the time averaging,
and foreground removal techniques (see Figure~\ref{fig:pipeline_compare}).

\begin{figure}
\centering
\includegraphics[width=.5\textwidth]{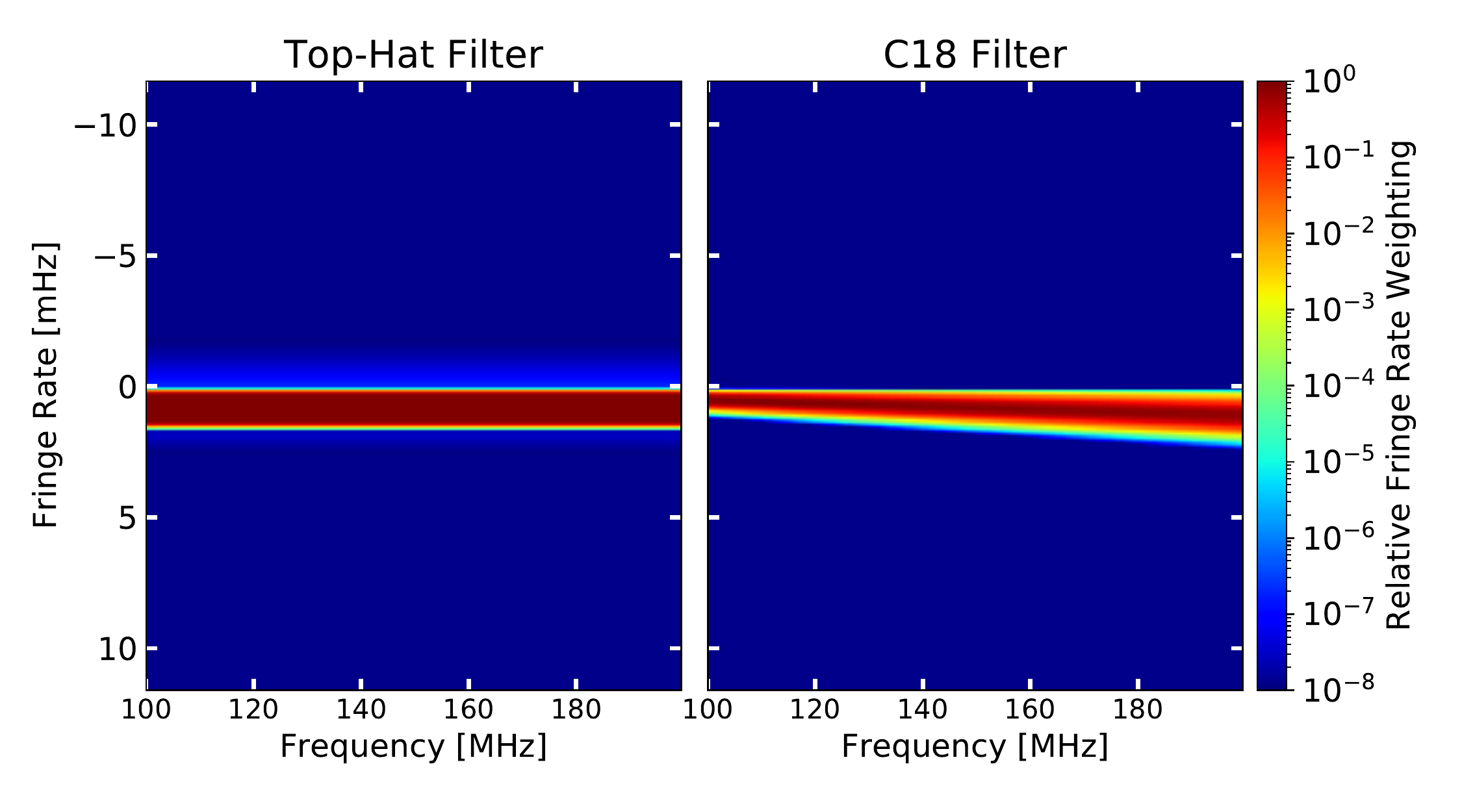}
\caption{A comparison of the Top-Hat fringe-rate filter (TH, left) and the filter used in \citetalias{cheng_et_al2018} (right) in the fringe-rate, frequency domain.
The \citetalias{cheng_et_al2018} filter
varies with frequency and this spectral
variation can cause additional structure
when performing a delay transform of the visibilities. In the interest of simplicity
in this analysis, we choose to perform time averaging with the
Top-Hat filter.
}\label{fig:FRF_response}
\end{figure}

\begin{figure}[tp]
\centering
\includegraphics[width=.5\textwidth]{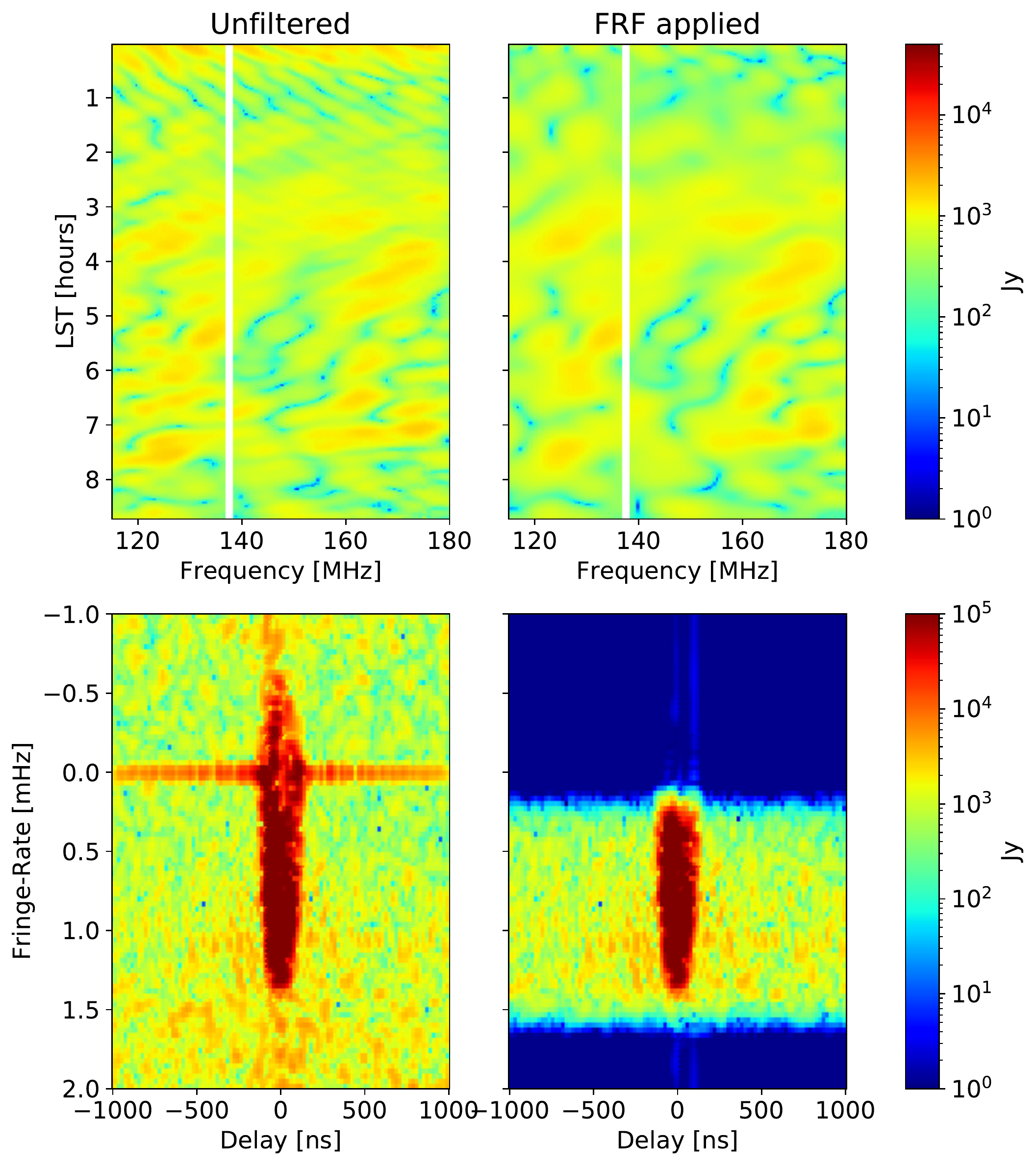}
\caption{\textbf{Top:} LST and frequency waterfalls of representative
baselines taken from the even LST binned set before (left)
and after (right) application of the Top Hat FRF.
The baseline illustrated is the antenna pairs (1,4). The application of the fringe-rate filter removes very fast
fringe modes but preserves the structure of sky-like
modes.
\textbf{Bottom:} The same baseline before (left) and after (right) the application of the FRF plotted in fringe-rate and delay space. The Fourier representation of the data illustrates the \textit{common mode} at fringe-rate=$ 0 $~mHz suppressed by the filter.}
\label{fig:waterfalls}
\end{figure}
\subsection{Time Averaging}\label{sec:frf}

The LST binned data have been \added{initially} averaged into $43$~s bins, a timescale
which is short compared to the $\approx$3500~s fringe coherence time of the 30\,m baselines (see Section 3.5 in \citetalias{ali_et_al2015}).
Here, as in past PAPER analyses, we choose to perform  \added{additional} time averaging by convolving the time stream with a windowing function.
This function is defined as a filter in fringe-rate space
(the Fourier dual to LST) which can be tuned to maximize sensitivity to sky-like modes and exclude slowly varying systematics.
\cite{parsons_backer2009} show that
a fringe-rate corresponds to sky-like rates of motion which map geometrically to
a great circle on the sky.
\cite{parsons_et_al2015} then shows that a fringe-rate filter (FRF) can be defined with weights corresponding
to the square root of the instrument's primary beam power squared and integrated along the line of constant fringe-rate. Applying an FRF with this weighting provides optimal thermal sensitivity in power spectrum
estimation.

Previous PAPER analyses have used variations on such a filter.  \citetalias{ali_et_al2015}
formed the beam-weighted filter, fit a Gaussian
in fringe-rate space, and then artificially increased
the width of the Gaussian
to provide easy parameterization across the PAPER
bandpass and decrease the effective time integration.
A similar Gaussian fit was
also used and discussed in \citetalias{cheng_et_al2018},
but the width of the fit was not increased in this
analysis.

However, as can be seen in the right hand side of the
top of Figure~\ref{fig:FRF_response}, this filter is frequency dependent.
In particular, the maximum fringe-rate range probed by a baseline increases linearly with frequency.
This spectral dependence may introduce additional structure during the delay transform; further investigation is needed to find the best approach for mitigating this effect.

Additionally, the use of these ``aggressive" fringe-rate filters have also
been shown to contribute to signal loss
\citepalias{cheng_et_al2018} especially when used in conjunction with quadratic power spectrum estimators.

 While the QE formalism  is not used in this work, as a simplification to avoid potential signal loss
and reduce contamination of high delay modes,
we adopt a Top-Hat filter
that weights all fringe-rates evenly across frequency, similar to the filter used in \citet{parsons_et_al2012b}.
The maximum fringe-rate passed by our filter is set
by the highest frequency
included in the data set; the lowest fringe-rate passed is chosen to exclude known common mode signals with zero fringe
rates. 

\added{This results in an effective integration time of $ \sim 940 $~s measured as the equivalent noise bandwidth of the windowing function. While the filter results in sub-optimal thermal sensitivity on the estimated power spectrum, it is designed to remove a \textit{common mode} signal observed in previous $ \PAPER $ analyses while providing a moderate increase in thermal sensitivity.
}

\subsubsection{Common Mode}\label{sec:common_mode}
Past PAPER analyses have noted
signals which vary on time scales
longer than would be expected from an ideal interferometer \citep{ali_et_al2015}.
Such common modes\footnote{Previously referred to as
	``crosstalk.'' These common mode signals may not
	necessarily result from signals observed in one antenna and leaked to another
	(a time delayed sky signal) but rather any time-independent signal which is observed by all antennas.}
are excluded here by setting the minimum fringe-rate included in the filter to $3.5\times10^{-5}$~Hz; this
excludes all modes with periods longer than $\sim$45 minutes.

Suppressing slowly or negatively fringing sources will suppress sources  with 
elevations at or below the south celestial
pole.  These modes are generally
low in the $\sim 45\arcdeg$ PAPER primary beam.
When applying this filter to our foreground simulation,
 the total simulated power is observed to decrease by $7.97\%$, as a result we apply a correction factor of 
 $ 1.086 $ to our power spectrum estimates and their uncertainties to account for 
 the associated signal loss.

Waterfall plots of a representative baseline
before and after the application of the
fringe-rate filer are shown in Figure~\ref{fig:waterfalls}.
The application of the fringe-rate filter removes very fast
fringe modes but preserves the structure of eastward moving sky-like modes. Also visible is the \textit{common mode} at fringe-rate=$ 0 $~mHz which is suppressed by the the application of the FRF. Without filtering, the \textit{common mode} would create a strong bias at high delay modes during power spectrum estimation.

\subsection{Foreground Removal}\label{sec:wida}
To mitigate foreground contamination during
power spectrum estimation, PAPER analyses have used
an wide-band iterative deconvolution algorithm (WIDA)
often referred to as a ``clean-like'' iterative
deconvolution algorithm. This algorithm relies on the
underlying mathematics of CLEAN as described in
\citet{hogbom1974} to remove delay components from
PAPER data inside of some range of delays.
This type of deconvolution and its specific
application to radio data is described in \citet{parsons_backer2009}.
The WIDA was used in \citet{parsons_et_al2012b,parsons_et_al2014,jacobs_et_al2014}, \citetalias{ali_et_al2015}, \citet{kerrigan_et_al2018}, and \citetalias{cheng_et_al2018}.

The use of this filtering technique has been omitted
from this analysis. While the technique should not
affect cosmological signals outside the user defined
range of delays to clean (\citealt{parsons_backer2009, parsons_et_al2012b, parsons_et_al2014}, and explored further in \citealt{kerrigan_et_al2018}),
recent works have also shown the use of this filter does not produce
statistically significant reduction of power at
high delay modes \citep{kerrigan_et_al2018}.
Since our analysis aims to focus on upper limits set
at high delay modes, we omit this step in the interest
of simplicity. Even without any attempt to remove foregrounds from the
visibility data, we find that our delay transform used to estimate
the cosmological power spectrum is not limited by the inherent
dynamic range of the transform.

\section{Redundancy of PAPER Baselines}\label{sec:redundancy}
\begin{figure}[tp]
	\centering
	\includegraphics[width=.5\textwidth]{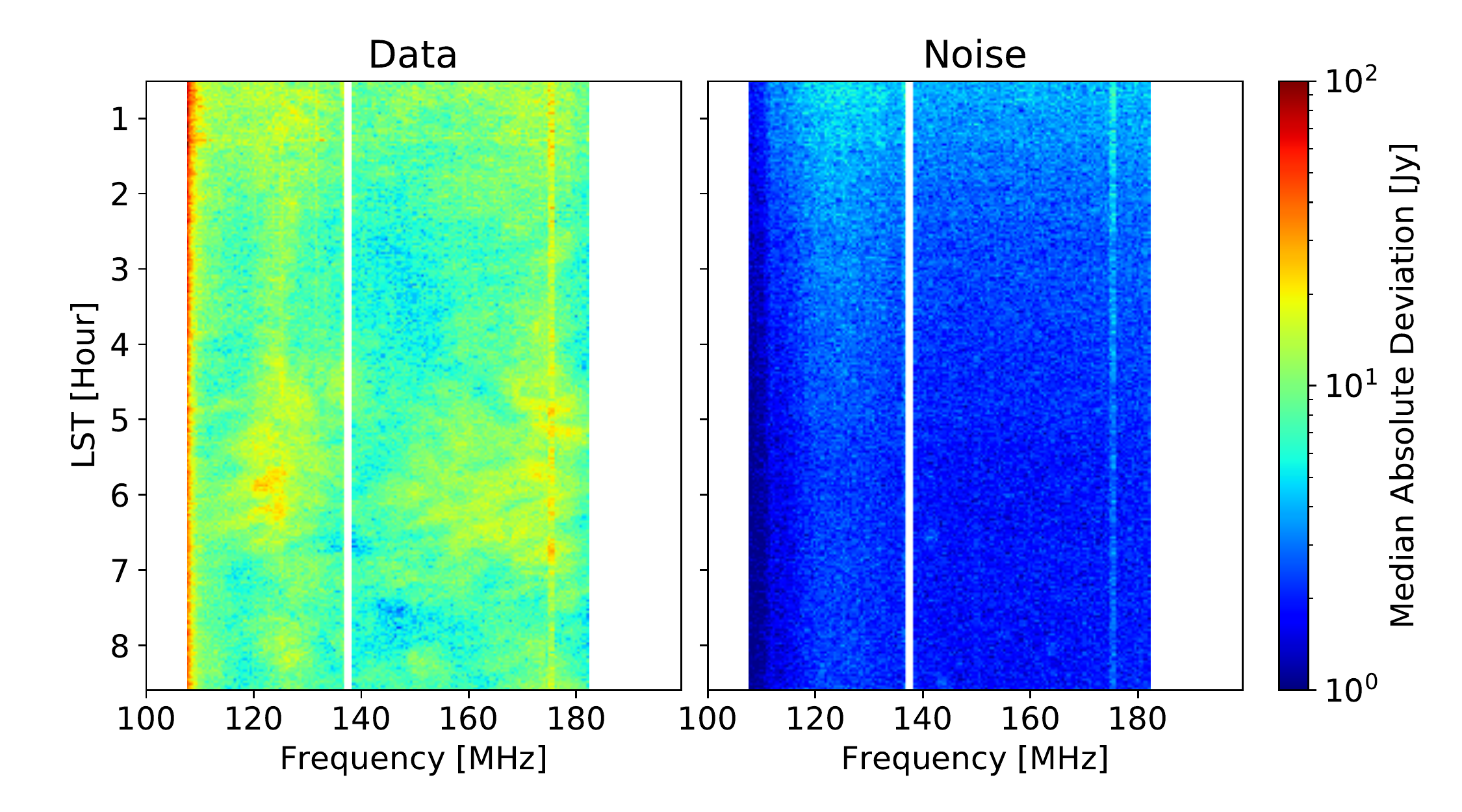}
	\caption{A representative Median Absolute Deviation (MAD) for both data (left)
		and noise simulation (right) computed for each time and frequency observed by
		PAPER in the LST range $00^{h}30^{m}00^{s} - 08^{h}36^{m}00^{s}$.
    The data shown here corresponds to
		strictly East-West baselines in Figure~\ref{fig:ant_pos}.
		For perfectly redundant sky measurements the individual
		baseline measurements will only differ by thermal noise.
		The large amplitude of deviations observed
		illustrates that there is a significant
		amount of non-redundant information in the data.}\label{fig:mad}
\end{figure}

\begin{figure*}[tp]
\centering
\includegraphics[width=\textwidth]{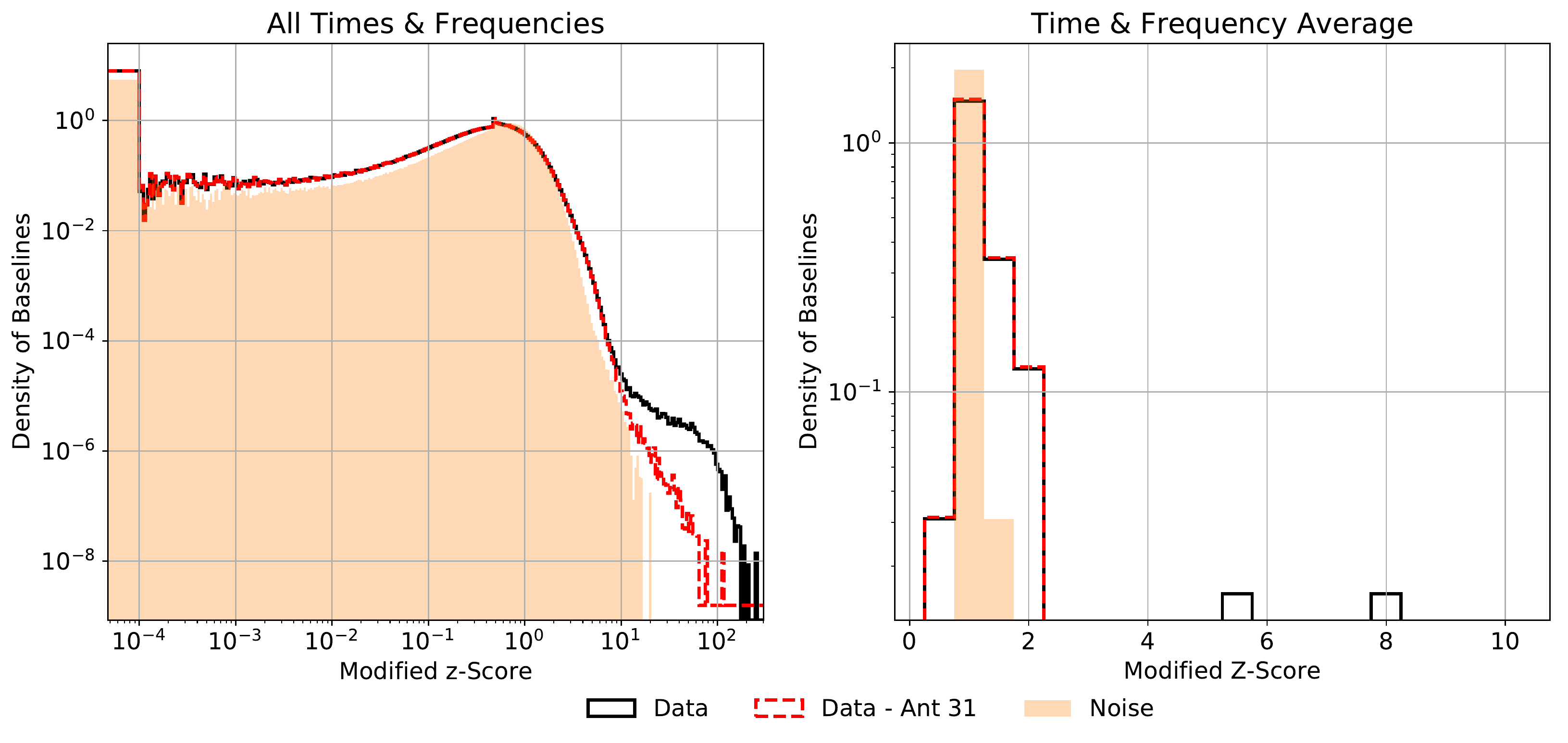}
\caption{A histogram of modified z-scores of data averaged in quadrature over LST day (even/odd) (black line) 
and input noise simulation also averaged over LST day (orange) before (left) and after (right) averaging in quadrature over frequencies and times. Also plotted is the 
distribution of z-scores after removing any identified outliers (dashed red).
The visual shoulder in the left hand plot near $ M_{z}\sim50 $ is evidence of 
non-redundant contributions larger than the fluctuations from thermal noise. 
To identify the contaminating baselines a 
quadrature averaged is performed of the frequency and time
axes to produce a single modified z-score per baseline.
The variance of the distribution of the noise is $ \sim40 $ times smaller than the distribution from the data. As such,
 performing a statistical cut based on the 
distribution of the noise simulation would result in 
removing $ \sim 85$~\% of all baselines. This is a result of the noise simulating a perfectly redundant set of baselines.
Therefore a visual inspection is necessary to identify 
potential outliers.
The two baselines $ (21, 31) $ and $ (31, 45) $ present as
obvious candidates for removal. Both baselines have 
modified z-scores greater than $ 4 $ and removing them 
is consistent with a cut at $ M_{z}=3.5 $ as suggested in \citet{Iglewicz_and_hoaglin}.
The removal of these two baselines also flags antenna $ 31 $ entirely from the analysis as it contributes only to these baselines.
}\label{fig:mod_z_score_avg}
\end{figure*}

Before estimating the power spectrum of the data,
we conduct statistical tests on the observations to determine the degree to which
the baselines are redundant.
The per baseline delay spectrum
estimation technique described in \citet{parsons_et_al2012b}
can be averaged across all baselines cross-multiples only for
perfectly redundant baselines\footnote{Or if the non-redundant component of an ensemble of baselines is described by a random variable with mean 0 (like Gaussian noise).}. While it is unrealistic to assume
the PAPER baselines are perfectly redundant, this analysis can
help identify extreme outliers which should not be used in
power spectrum estimation.

As discussed in Section \ref{sec:data_selection},
the 8 x 8 antenna configuration used in the PAPER-64
deployment was chosen to increase sensitivity on baselines
with many redundant observations. Each of the three baseline vectors are
sampled many times across the grid-like array. Rather than average
baselines together (as was done in previous PAPER analyses for computational simplicity),
we cross multiply all redundant pairs and then bootstrap average for
an estimate of variance. This is described in more detail in Section \ref{sec:bootstrap}.

A first test of the array's redundancy is to compare the measured variation between baselines
with that expected due to thermal noise, using the input noise simulation discussed in Section~\ref{sec:noise_sim}.

As a measure of variance between baselines, we take the Median Absolute Deviation
(MAD) of the visibility amplitude across redundant baselines
for each frequency and time, defined as
\begin{equation}
\text{MAD}(t,\nu) = median\left(\left| \left|V_{i,j}(t,\nu)\right|  - median \left( \left|V(t,\nu)\right|\right) \right|\right)\label{eqn:mad}
\end{equation}
where the median visibility amplitude is taken
at each time and frequency across the redundant baseline group.

The MAD for both data and our noise simulation is shown in  Figure~\ref{fig:mad}.
For perfectly redundant sky measurements the individual
baseline measurements will only differ by thermal noise.
Some frequency, time pairs have a MAD consistent with thermal noise,
however the larger deviations observed at other frequencies and times
illustrates a significant amount of non-redundant information in the data.

We then use the MAD to estimate the significance of each baseline's
deviation from the median baseline measurement using the modified z-score ($ M_{z}(t,\nu) $) defined as
\begin{equation}
M_{z}(t,\nu) = 0.6745\frac{\left|V_{i,j}(t,\nu) - median \left(V(t,\nu)
	\right)\right|}{MAD} \label{eqn:zscore}
\end{equation}
which can be thought of as the number of ``sigma''
away from the median each data point is.
The $ 0.6745 $ scaling factor is introduced to normalize the
modified z-score for a large number of samples \citep{Iglewicz_and_hoaglin}.

These scores, $ M_{z} $ are calculated for each set of LST binned data (even and odd). In order to provide an estimate of a single $ M_{z} $ for every baseline, the modified z-scores are initially averaged in quadrature of the LST day dimension.
The histograms of these modified z-scores averaged in quadrature  over 
LST day for both the input data and noise simulation are shown in left had side of Figure~\ref{fig:mod_z_score_avg}.
A quadrature average is chosen to identify absolute outliers
opposed to an unweighted averaged where a hypothetical 
baseline with an even distribution of positive and negative outliers could average to zero.

The distribution of modified z-scores for all frequencies
and times illustrates a significant of non-redundant signal
beyond the contributions from thermal fluctuations. 
To better identify the baselines (or antennas) contributing
to this non-redundant information,
a quadrature average is performed over the frequency and
time dimensions for all baselines and the resulting 
distribution is shown in the right hand side of Figure~\ref{fig:mod_z_score_avg}.

Since the noise simulation is a model of perfect redundancy,
the quadrature averaging produces a very narrow distribution
centered near $1$.
As a result, it is impossible to remove only a small number
of outlier baselines (or antennas) using a cut based on the
distribution of scores from the noise simulation.
The variance of the distribution of the noise is $ \sim40 $ times smaller than the distribution from the data. As such,
performing a statistical cut based on the 
distribution of the noise simulation would result in 
removing $ \sim 85$~\% of all baselines. 
This redundancy analysis is aimed to only remove the worse
antenna (or two) from the analysis, not drastically reduce
the number of input baselines.
Therefore a visual analysis of the distribution of modified z-scores is necessary to identify potential outliers.

The two baselines $ (21, 31) $ and $ (31, 45) $ present as
obvious candidates for removal. Both baselines have 
modified z-scores greater than $ 4 $ and removing them 
is consistent with a cut at $ M_{z}=3.5 $ as suggested in \citet{Iglewicz_and_hoaglin}.
The removal of these two baselines also flags antenna $ 31 $ entirely from the analysis as it  contributes only to these baselines.
The distribution of modified z-scores without this outlier
antenna is also plotted in Figure~\ref{fig:mod_z_score_avg}.

Although no other baselines qualify as outliers,
the difference in distributions between the data and noise 
simulation indicates an amount of non-redundancy
significantly inconsistent with thermal noise fluctuations from the baselines in this analysis and may affect
the interpretation of our final
power spectrum estimates.

\section{Power Spectrum Estimation}\label{sec:OQE}

\added{Our analysis pipeline uses a delay-based power spectrum estimation technique first developed in \citet{parsons_et_al2012b}. 
The highly redundant baseline configuration in $ \PAPER $ provides high thermal sensitivity on a small subset of short ($ \sim 30 $~m) baselines by observing repeated samples of the same sky signals with independent noise \citep{parsons_et_al2012a}. Also, the fringe spacing corresponding to the baselines and observing frequencies probe a single spatial fluctuation scale ($ k_{\perp} $-mode) as a function of frequency. By Fourier transforming along the frequency axis into delay space, foregrounds are expected to be constrained to an area bound by the maximum geometric delay of the chosen baseline \citep{parsons_et_al2012b}. Additionally, \citet{kerrigan_et_al2018} shows that an application of foreground subtraction applied to delay-based power spectrum estimators only affects delay modes just outside of the geometric delay limit of a baseline. As such, thermal sensitivity at delay modes larger than the maximum geometric delay of a baseline should be unaffected whether foregrounds are subtracted or not. The high thermal sensitivity and constrained foreground in delay-space make $ \PAPER $ well suited for a delay-based power spectrum estimation.}

The power spectrum of the 21cm emission can be estimated
directly from interferometric visibilities following
\citet{parsons_et_al2012a} and \citet{parsons_et_al2014}.
\begin{equation}
P(k_{\parallel}, k_{\perp}) =\left(\frac{\lambda^{2}}{2k_{B}} \right)^{2} \frac{X^{2}Y}{\Omega_{eff}B_{pp}}
\left\langle \tilde{V}^{\star}_{i}(\tau,t)\tilde{V}_{j}(\tau,t) \right\rangle_{i\neq j,LST} \label{eqn:pspec}
\end{equation}
where $ \lambda $ is the observed wavelength,
$X^{2}Y$ converts from interferometric units to
cosmological units, $ k_{B} $ is the Boltzmann constant,
$ \Omega_{eff} $ is the effective area of the primary beam depending on the units of the input visibility \citep{parsons_et_al2014,parsons_et_al2015}, $ B_{pp} =\int d\nu |\phi(\nu)|^{2} $
is the effective bandwidth of the power spectrum estimation where $ \phi(\nu) $ is the spectral taper function used during Fourier transform, and $ \tilde{V}_{i}(\tau, t) $ is the delay
transformed visibility observed by baseline $ i $.
This formula assumes the baselines over which the delay
transform is taken have a minimal change in length over the
bandwidth of the transform. This allows for a one to one correspondence 
between the delay modes as a
function of $ \tau $ and the cosmological modes,
$k_{\parallel} $ \citep{liu_et_al2014a}.

The power spectrum is estimated
by selecting subsets of available bandwidth,
weighting by a tapering function ($ \phi(\nu) $) to improve dynamic range,
 delay transforming visibilities with an FFT,
  cross-multiplying different baseline pairs, and then
  bootstrap-averaging cross-multiplication pairs.
  Foreground leakage in the FFT is minimized with a
  Blackman-Harris (BH) tapering function before the Fourier transform over frequency.
  The BH window does induce a correlation between directly adjacent
  Fourier modes, however, and the resulting bandwidth/redshift range sampled by each power spectrum window
   is effectively halved for each redshift band.

These steps are implemented with the
publicly available
simpleDS\footnote{\url{github.com/RadioAstronomySoftwareGroup/simpleDS}}
python package.
This package and analysis pipeline have been developed
 specifically to provide a simple alternate analysis to other
 pipelines that take more aggressive strategies with regard to weighting and foreground removal.

\begin{deluxetable*}{lchccccccc}[tp]
\centering
%\tablewidth{\textwidth}
\tablecaption{PAPER-64 Theoretical Noise Estimate Values}
\tablecolumns{10}
\tablehead{
    \multirow{2}{*}{Term} &
	\multirow{2}{*}{Description} &
	\multirow{2}{*}{Formula} &
	\multicolumn{6}{c}{Value in Redshift bin} &
	\multirow{2}{*}{Units} \\
    &
    &
    &
    \colhead{$10.87$} & 
    \colhead{$9.93$} & 
    \colhead{$8.68$} & 
    \colhead{$8.37$} &
    \colhead{$8.13$} & 
    \colhead{$7.48$} & 
}
\startdata
$X^{2}Y$ & \parbox{6cm}{\centering Conversion from interferometric $(u,v,\eta)$ to cosmological $(k_{\perp,x}, k_{\perp,y}, k_{\parallel})^{\footnotesize  a}$} & $ \frac{c(1+z)^{2}D_{c, \perp}(z)^{2}}{\nu_{21} H_{0} E(z)} $ &  578.77 & 533.36& 471.06 & 454.90 & 442.58 & 408.52 & $\frac{\text{Mpc}^{3}}{h^{3} \text{sr Hz} }$  \\
$\Omega_{eff}$ & Effective beam area$^{\footnotesize b}$& $\frac{\Omega_{p}^{2}}{\Omega_{pp}} $ & $1.645$ & $1.664$ & $1.489$ & $1.487$ & $1.496$ & $1.580$ & sr \\
$T_{sys}$ & System temperature & $180\left(\frac{\nu\ [GHz] }{.18} \right)^{-2.55}\ +\ T_{rcvr}$ & $653.37$ & $556.33$ & $446.75$ & $422.31$ & $404.69$ & $360.30$ &  K\\
$T_{rcvr}$ & Receiver Temperature & & \multicolumn{6}{c}{\dotfill$144$\dotfill} & K \\
$N_{lst}$ & Number of effective LST samples & &  \multicolumn{6}{c}{\dotfill$31$\dotfill} &  \\
$N_{sep}$ & Number of independent baseline types$^{\footnotesize c}$ & & \multicolumn{6}{c}{\dotfill$3$\dotfill} &  \\
$t_{int}$ & Integration time of LST sample$^{\footnotesize d}$ & & \multicolumn{6}{c}{\dotfill$938$\dotfill} & s \\
$N_{days}$ & Number of effective days used in LST-binning & &  $27.63$ & $27.81$ & $28.07$ & $28.33$ & $28.44$ & $28.79$ &  \\
$N_{pols}$ & Number of polarizations combined in analysis & & \multicolumn{6}{c}{\dotfill$2$\dotfill}& \\
$N_{bls}$ & Number of effective baselines & & \multicolumn{6}{c}{\dotfill$47$\dotfill} &
%\dots & \makebox[4.95cm]{\dotfill} one N-S  separation & $40$  & \\
%\dots & \makebox[4.1cm]{\dotfill} minus one N-S separation & $40$  &
\enddata
\tablenotetext{a}{This value is also a function of the assumed
	background cosmology. See \citet{furlanetto_et_al2006} and \citet{liu_et_al2014a} for more information.}
\tablenotetext{b}{The effective beam area is influenced by the
	choice of fringe-rate filter applied \citep{parsons_et_al2015}.
	The computation for this value is also found in Appendix B of \citet{parsons_et_al2014}.}
\tablenotetext{c}{The ``sep'' subscript refers to the separation
	of antennas on the PAPER-64 grid. These separations are what define the different baseline types described in Figure~\ref{fig:ant_pos}.}
\tablenotetext{d}{This value is computed as the Equivalent Noise
	Bandwidth (ENBW) of the FRF applied to the data. See \citetalias{cheng_et_al2018} and \citet{parsons_et_al2015} for more information.}
\label{tab:noise_terms}
\end{deluxetable*}
%\begin{comment}

\subsection{Power Spectrum Uncertainties}
In this section, we present several different methods for estimating the
uncertainties on our power spectrum estimates.  Combined, these alternative
approaches help provide a consistent picture of the uncertainties on our results.

\subsubsection{Bootstrapped Variance}
\label{sec:bootstrap}
Power spectrum errors can come from thermal, instrumental, and terrestrial (RFI) sources. Biases
and additional variance can also be unintentionally introduced
in analysis steps (e.g. calibration, time-averaging).
Those with a known covariance (like thermal noise) can be propagated through the
data processing and power spectrum estimation steps into an analytically estimated error bar. The other
sources are harder to estimate from first principles.  However, the total variance of the data --- independent of the exact source of error --- can be estimated by bootstrapping: estimating the power spectrum
from subsets of data and then calculating the variance of these estimates. In the
redundant PAPER array, the axis most amenable to bootstrapping is the
selection of baseline pairs which are cross multiplied to get a power spectrum.

We provide an overview of the bootstrapping technique used in this work below. This method incorporates
the bootstrapping revisions described in more detail in Section~3.2.2 of \citetalias{cheng_et_al2018}.
Specifically, we perform the power spectrum estimation by cross
multiplying all pairs of baselines within a redundant set and \emph{between} the two even and odd
LST binned data sets described in Section \ref{sec:data}.  These cross
multiplications are then randomly sampled with replacement and then averaged
over all cross-multiple products resulting in a single waterfall of power spectra. An average is then taken across the LST axis
to form a single power spectrum versus delay.

We repeat this process by selecting different baseline
cross-multiplications to find new realizations of the power spectrum. The variance
of these bootstrap samples is interpreted as
the uncertainty in the power spectrum estimate.
This bootstrap estimation is designed to probe the underlying
distribution of allowed values given our observed values
\citep{efron_tibshirani1994,andrae2010}.

\subsubsection{Thermal Variance}
\citet{liu_et_al2014a,liu_et_al2014b} show that when
 estimating the power spectrum in
the regime $k_{\parallel}\gg k_{\perp}$,
the delay axis (the Fourier dual to frequency) can
(to a good approximation) be re-interpreted as
the cosmological $k_\parallel$ axis.
Under this assumption, to provide a theoretical estimate of the
thermal variance, we use the expected noise power derived in \citep{parsons_et_al2012a} and applied in \citet{pober_et_al2013, pober_et_al2014} and \citetalias{cheng_et_al2018}.
\begin{equation}
P_{N}(k) = \frac{X^{2}Y \Omega_{eff} T_{sys}^{2}}{t_{int} N_{days} N_{bls} N_{pols}\sqrt{2 N_{lst} N_{sep}}} \label{eqn:theory_noise}
\end{equation}
where $X^{2}Y$ converts from interferometric units to cosmological
units, $T_{sys}$ is the system temperature, $\Omega_{eff}$ is the
effective size of the primary beam in steradians \citep{parsons_et_al2014}, $N_{lst}$ is the
number of independent LST samples,
$N_{pols}$ is the number of polarizations used in the analysis,
$t_{int}$ is the integration time of an LST sample,
$N_{days}$ is the effective number of days used in LST
binning, $N_{bls}$ is the effective number of baselines combined,
and $N_{sep}$ is the number of independent baseline types.
See \citetalias{cheng_et_al2018} for a thorough definition of 
all the terms in this thermal noise estimate.

This estimate assumes the number of times each LST is observed the same number of times across the full course of the season, when in practice LSTs were observed between 5 and 60 times (a consequence of only observing at night with a drift-scanning telescope).
These counts are tabulated during the LST binning process.  %Assuming Gaussian samples
If the noise is constant from night to night, an effective $N_{days}$ can be calculated by averaging the inverse sum of squares over the sidereal period
as described in \cite{jacobs_et_al2016}. The observations here yield an effective integration length of varying between $ 27 $ and $ 29 $~days depending on the redshift bin.

As an aid to future repeatability, the values used here
are listed in Table~\ref{tab:noise_terms} and the calculation is documented as a python module called \code{21CMSENSE\_CALC} available at \url{github.com/dannyjacobs/21cmsense_calc}.

Equation~\ref{eqn:theory_noise} is an analytic form that serves as a useful ``sanity check'' on the expected noise levels, but is not expected to be highly accurate in the absence of simulations to calibrate its terms.  Simulation of $ T_{sys} $ through the power spectrum pipeline (the nose input described in Section~\ref{sec:noise_sim}) is likely to be the most robust estimate of thermal noise errors.

\subsubsection{Foreground Error Bars}
\label{sec:FG_err}
The propagation of the thermal error above does not fully capture the
variance expected on modes with significant non-noise like power (i.e. foregrounds).
To demonstrate this fact, let us assume each visibility, $ \tilde{V}_{i} = s + n_{i} $,
to be the sum of a signal component,
$ s $,  and some noise term, $ n_{i} $.
Assuming the signal component is
constant across all baselines and
$ n_{i} \sim \mathcal{CN}(0,\sqrt{P_{N}}) $ is independent on
every baseline $ i $, we show in Appendix~\ref{appendix:var}
that the variance of each power spectrum cross-multiple can be
written as
\begin{equation}
\sigma_{P(k)}^{2} \equiv Var(P(k)) = 2P_{s}(k)P_{N} + P_{N}^{2} \label{eqn:prop_error}
\end{equation}
where $ P_{s}(k) $ is the true power spectrum of the sky
signal and $ P_{N} $ is the noise power spectrum from
Equation~\ref{eqn:theory_noise}.

 The signal-noise cross-term in Equation \ref{eqn:prop_error}
will dominate delay/$k$-modes inside
the horizon where the expected foreground signal exceeds
thermal noise levels. At the highest delay/$k$-modes, the
 uncertainty will be dominated by the thermal
variance.

Here we use the simulated PRISim observation as a rough
estimate of $ P_{s}(k) $, with the simulation scaled to match the data on average
across the entire band
\begin{equation}
P_s(k) = g^{2} P_{\text{PRISim}}(k) \label{eqn:ps_proxy}
 \end{equation}
where $ g $ is the model scale factor computed in Section~\ref{sec:sims}.

\begin{figure*}[tp]
\centering
\includegraphics[width=\textwidth]{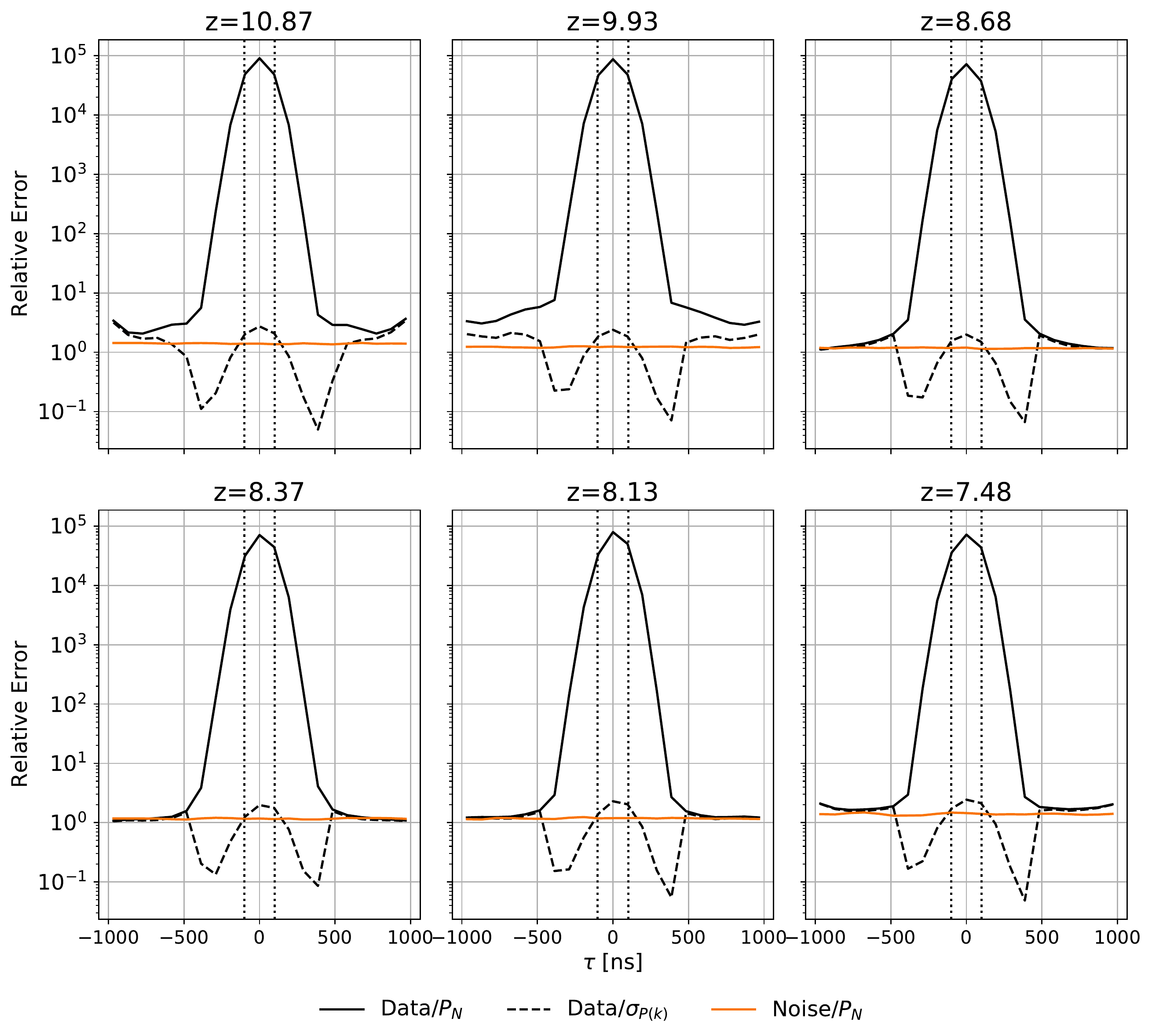}
\caption{The ratio of the bootstrap error bars of both data and noise to estimates of the predicted uncertainties for each redshift bin.
Panels are ordered such that redshift increases towards the upper left.
A ratio helps to compare different estimates of power spectrum error bars together.
Bootstrapped errors of simulated noise (orange) over $ P_N(k) $ (Equation~\ref{eqn:theory_noise})  are very close to
unity ratio, an important consistency check.
 The ratio of data variance to $ P_N(k) $ (solid black line) is nearly unity like at high delay but is $10^4 \times$ higher where the simulated foregrounds dominate (refer to Figure~\ref{fig:pspec_vs_sim} to identify these regions). Accounting for the foreground dependent term in the theoretical error bar in the ratio denominator (dashed black line, $ \sigma_{P(K)} $; Equation~\ref{eqn:prop_error}), agreement closes by three orders of magnitude with the largest discrepancies now only a factor of 10 in the modes with the weakest foreground amplitude.
 This  order of magnitude of disagreement outside the horizon at
$ |\tau| > 100 $~ns may be the result of an incomplete sky model.}\label{fig:errorbar_ratio}
\end{figure*}

\subsubsection{Comparison of Power Spectrum Uncertainties}\label{pspec_err_compar}
As an internal consistency check, we compare the sizes
of the bootstrapped uncertainties to the analytical thermal noise
and simulated foreground uncertainties.
This comparison is made by taking the ratios of each type of uncertainty, which is plotted in Figure~\ref{fig:errorbar_ratio}.

As a basic test we see that the bootstrap variation of the external noise simulation (plotted in orange) is never further than a factor of 0.7
away from the theoretical prediction of purely thermal noise.
Considering the $10^9$ dynamic range spanned by power spectrum values, and remembering that the theoretical error bar includes several approximations, a 30\% worst-case difference is within expectations.

Bootstrapped errorbars of the data  are significantly larger than the purely thermal variance, sometimes reaching $10^5 \times$ larger in the horizon and nearly $5\times$ thermal noise at the highest redshifts in what, according to the simulations, should be noise dominated bins. In general, the overlarge error bars seemingly trace out all areas where the mean power spectrum itself manifests a notable excess.

However, accounting for the PRISim simulated foreground terms in the expected variance in the denominator of this ratio, agreement increases by orders of magnitude (the dashed curve), with the largest discrepancies now only a factor of $\sim10$. The remaining disagreement is concentrated in the modes where simulations show the weakest foreground amplitude that is still detectable above the noise.  This
remaining discrepancy at
$ |\tau| \sim 400 $~ns may be sourced from an incomplete sky model.

The addition of simulated foreground power to the noise calculation accounts for the largest discrepancies
in error estimation; however it does not decrease the discrepancy at high delays. At redshifts 8.68 and below
the difference between the calculated error estimates, the boostrapped errors, and the noise simulation all generally agree.  However, in
the two highest redshift bins the bootstrapped error estimate remains roughly $2$ to $5\times$ larger.
In all subsequent
analysis we include all three noise estimates
as useful comparisons.

\begin{figure*}[tp]
\centering
\includegraphics[width=\textwidth]{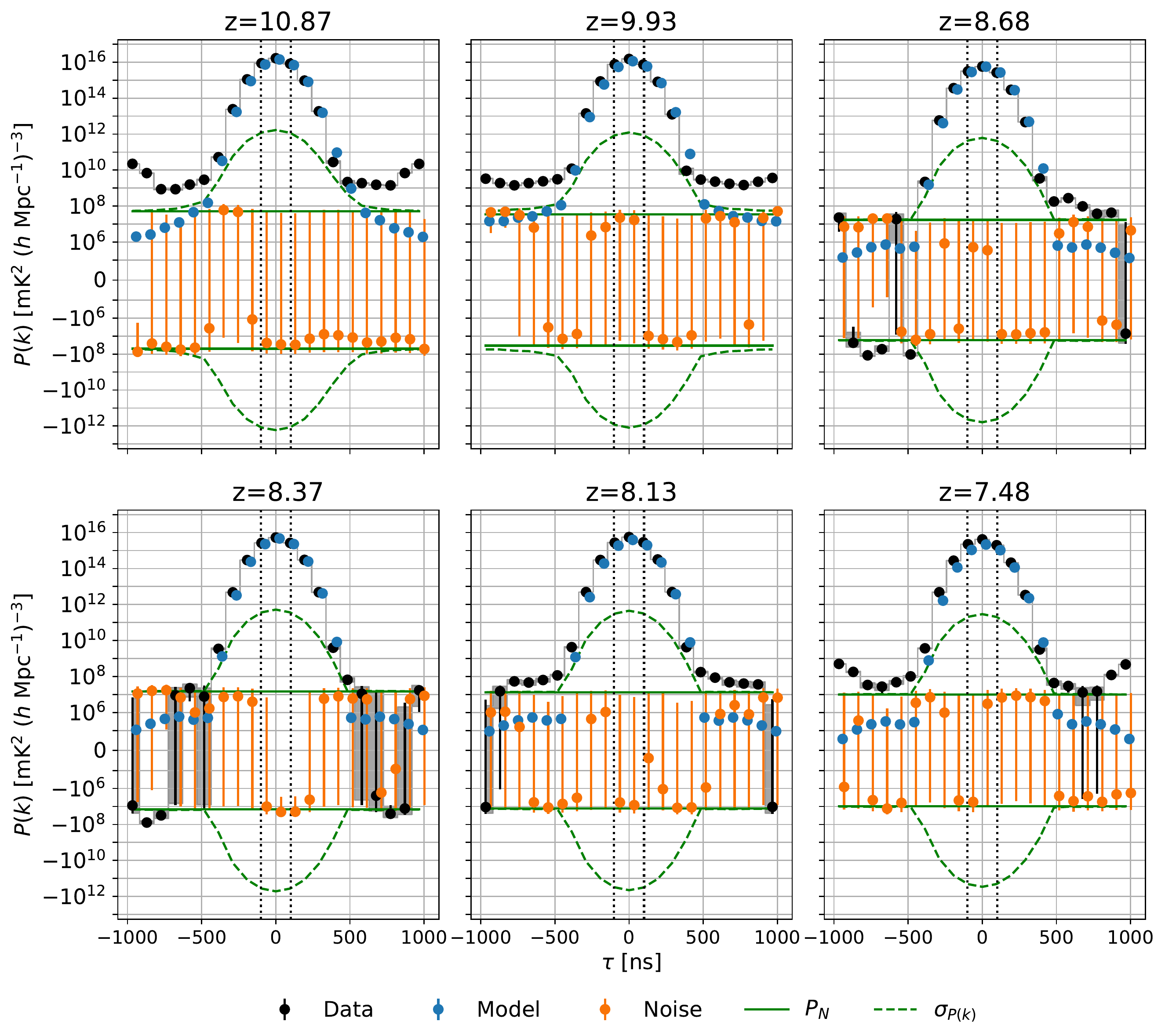}
\caption{Power spectrum estimates computed for
the observed data (black), simulated noise (orange), and simulated observation (blue).
Error bars on points are the bootstrapped uncertainty.
The solid green line indicates the theoretical thermal
noise estimate for each redshift bin, and the dashed
green line includes foreground error from Equation~\ref{eqn:prop_error}.
Grey shaded regions  are the foreground dependent uncertainties plotted around each data point.
The vertical black dotted lines indicate the horizon/wedge/light travel time for a 30~m
baseline.
As shown in Figure \ref{fig:errorbar_ratio}, the simulated noise is consistent
the theoretical thermal noise predictions.
At delay $\tau = 0$\,ns, both the the data and PRISim simulation
show good agreement in the total power observed;
generally, the power at all delays inside the horizon agrees between the two simulations within a factor of $ \sim5 $. The simulated data set also shows some
power leakage outside the horizon,
consistent with the power observed by $ \PAPER $ out to $ \approx 400$\,ns.  The PAPER data also show numerous statistically significant detections beyond 400\,ns, however, which are not predicted by the PRISim simulation.
To investigate the origin of these signals, multiple jackknives and null-tests are performed as described in Section~\ref{sec:nulls}.}
\label{fig:pspec_vs_sim}
\end{figure*}

\section{Multi-Redshift Power Spectrum Results}\label{sec:pspec_results}
Figure~\ref{fig:pspec_vs_sim} shows the delay power spectrum estimates for all
three of our principal products: the observed data (black),
the PRISim simulated observation (blue), and the noise only simulation (orange).
Within delay modes between
$ \sim\pm 400 $~ns, both the observed and simulated data illustrate
similar shapes. This suggests that the statistically significant detections of power
observed in PAPER immediately outside the horizon limits
are consistent with foreground signals (as suggested by the study of foreground subtraction applied to PAPER data in \citealt{kerrigan_et_al2018}).
At larger delays, however, the PAPER power spectra are a mix of statistically significant detections and
null results.  The most statistically significant detections at high delays are seen to occur at the lowest frequencies.

\begin{figure*}[tb]
\centering
\includegraphics[width=\textwidth]{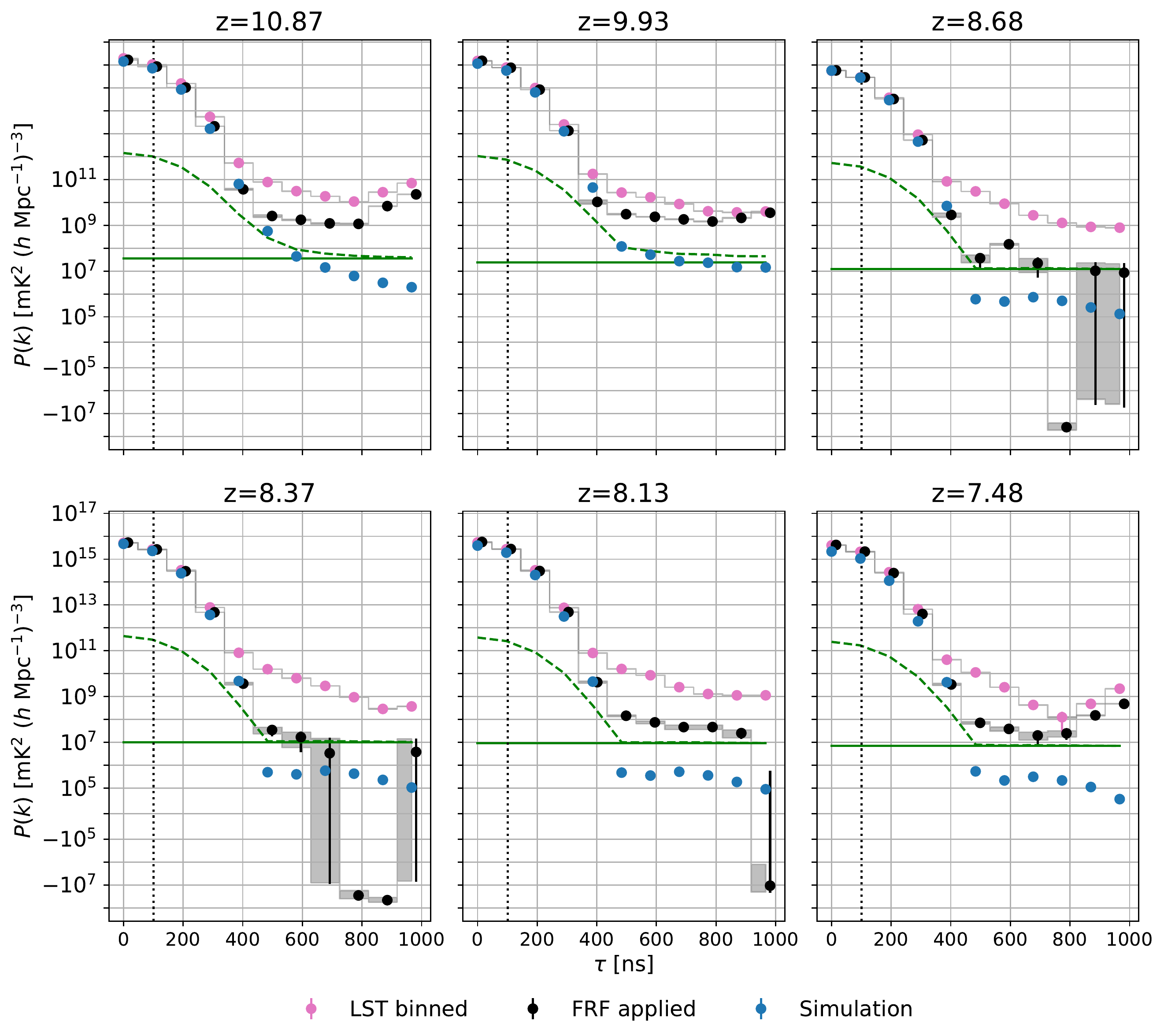}
\caption{The estimated power spectrum value before (purple) and after (black) application of the fringe-rate filter. The simulated data points (blue) have also been filtered with the FRF (the same as in Figure~\ref{fig:pspec_vs_sim}).  All other points and lines are the same as Figure~\ref{fig:pspec_vs_sim}.
While the application of the fringe-rate filter provides some improvement in thermal noise, it also provides suppression to the highly significant detections at delays $ |\tau| > 400 $~ns. These detections are inconsistent with the expected leakage from the simulated foreground signal (blue) and are signatures of the \textit{common mode} described in Section~\ref{sec:common_mode}.}\label{fig:pspec_vs_frf}
\end{figure*}

\begin{figure*}[tp]
\centering
\includegraphics[width=\textwidth]{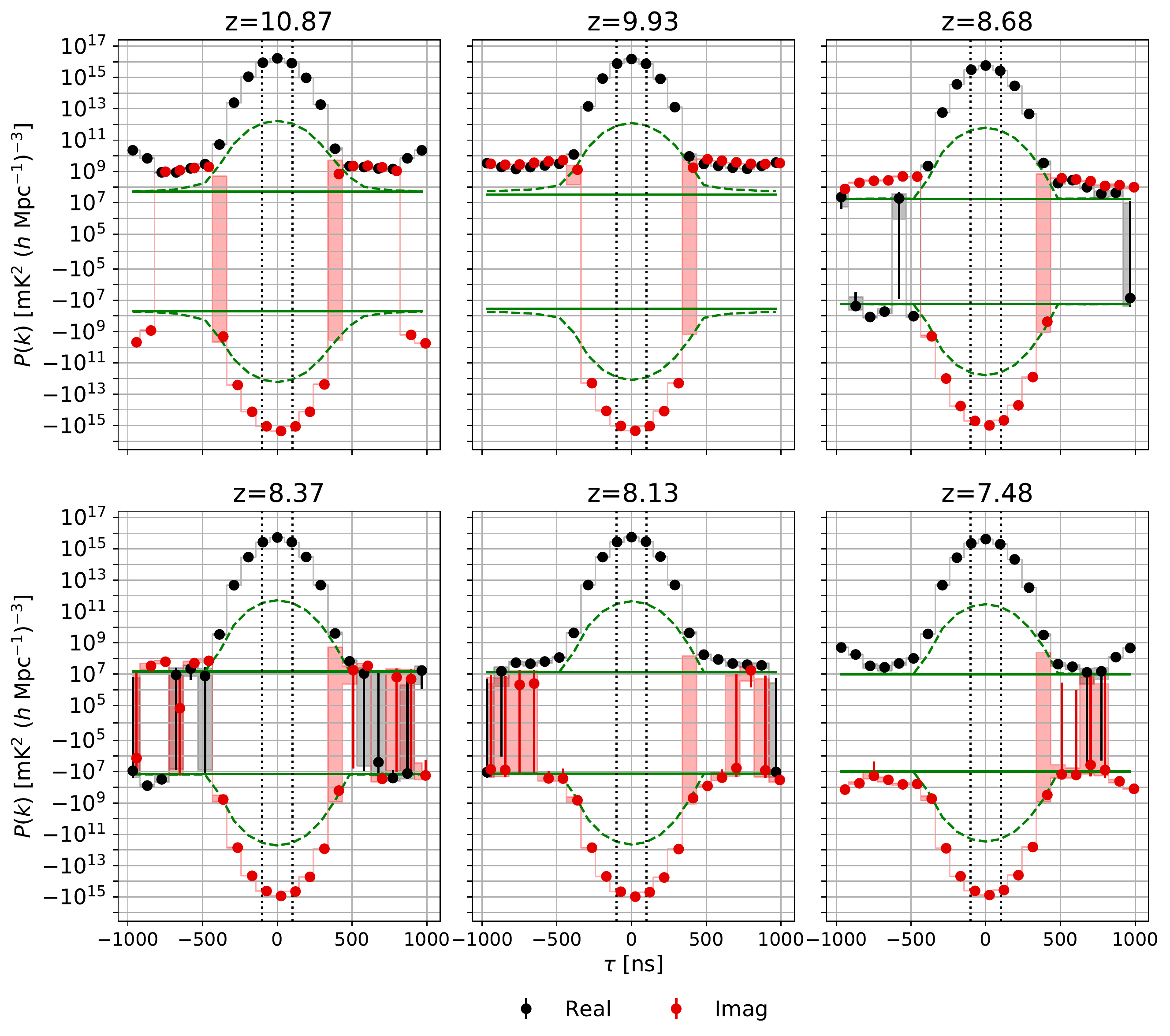}
\caption{The real (black) and imaginary (red) components of the
power spectrum of PAPER data. The red shaded region is the 
foreground dependent theoretical errorbar drawn around the 
imaginary components; all other lines are the same as in Figure~\ref{fig:pspec_vs_sim}. There are statistically significant imaginary components at
$ |\tau| < 400 $~ns, generally at a power level
which is $ \sim 20\% $ of the real components at the same delay.
All the detections in this region are also biased to
negative power levels.
This may result from non-redundancies in calibration or baseline orientation. At delay modes $ |\tau| > 400 $~ns, the imaginary
component of the power spectrum displays comparable power to the 
real part. This is especially prominent in, but not isolated to, the two highest redshift bins. 
The statistically significant imaginary power is indicative of some non-redundant information 
during power spectrum estimation, systematic biases introduced during data analysis or calibration, 
or residual contaminants like improperly flagged RFI.
}
\label{fig:real_imag}
\end{figure*}

\subsection{Evaluation of Fringe-Rate Filter}
The effectiveness of the fringe-rate filter in down 
weighting contaminating delay modes, can be evaluated 
after performing power spectrum estimation.

The power spectrum estimates before and after the application of the fringe-rate filter are shown in Figure~\ref{fig:pspec_vs_frf}.
While the application of the fringe-rate filter provides some improvement in thermal noise, it also provides suppression to the highly significant detections at delays $ |\tau| > 400 $~ns. These detections are inconsistent with the expected leakage from the simulated foreground signal (also filtered with the FRF)  and are signatures of the \textit{common mode} described in Section~\ref{sec:common_mode} and also visible in Figure~\ref{fig:waterfalls}.

Even for less aggressive filters than the ones used in \citetalias{ali_et_al2015} and \citetalias{cheng_et_al2018},
filtering can significantly reduce systematic 
contamination during the delay transformation.
The choice of the shape of filters is contingent on the 
acceptable amount of signal loss. As described in 
Section~\ref{sec:common_mode},
when applying this filter to our foreground simulation,
the total simulated power is observed to decrease by $7.97\%$, as a result we apply a correction factor of 
$ 1.086 $ to our power spectrum estimates to account for 
the associated signal loss.

\subsection{Investigation of High Delay Detections}

In this section, we present several analyses designed to help determine the cause of the remaining statistically significant detections at high delays seen in the PAPER observations.

\subsubsection{The Imaginary Power}

The power spectrum is computed by cross-multiplying
different baseline pairs within redundant groups.
Ideally, this cross-multiplication of complex-valued
delay spectra will result in any sky-like power being confined to the real part in the power spectrum, leaving the imaginary part dominated by noise.
However effects can leak real sky power into the
imaginary part of the power spectrum.
A perfectly calibrated array with non-redundant baselines
--- for example, with slightly different antenna positions ---
will cause two nominally ``redundant'' baselines to have slightly different phases.
The imaginary parts of these cross multiplied visibilities will therefore not cancel out, and non-zero power will be seen in the imaginary component of the power spectrum estimate.
The same effect would come from a perfectly redundant but imperfectly calibrated array.
It is also important to note that because of the foreground dependent error bars derived in Section~\ref{sec:FG_err},
imaginary power should increase at low delay, though continue to be consistent with zero.
In a sense, the amount of statistically significant power in the imaginary component of the power spectrum, compared to power in the real part,
is a measure of the net redundancy and calibration quality of the array.

A comparison of the real and imaginary parts of the power spectrum
is shown in Figure~\ref{fig:real_imag}.
The statistically significant imaginary components at
$ |\tau| < 400 $~ns are generally at a power level
which is $ \sim 20\% $ of the real components at the same delay.
All the detections in this region are also biased to
negative power levels. This may result from non-redundancies in calibration or baseline orientation.

At delay modes $ |\tau| > 400 $~ns, the imaginary
component of the power spectrum displays comparable power to the 
real part. This is especially prominent in two highest redshift bins, but is observable across the entire band.
The disagreement between the imaginary component and the foregrounds dependent thermal uncertainty is indicative of some non-redundant information,
systematic biases introduced by data analysis or calibration steps, or residual contaminants like improperly flagged RFI

\begin{figure*}[tp]
\centering
\includegraphics[width=\textwidth]{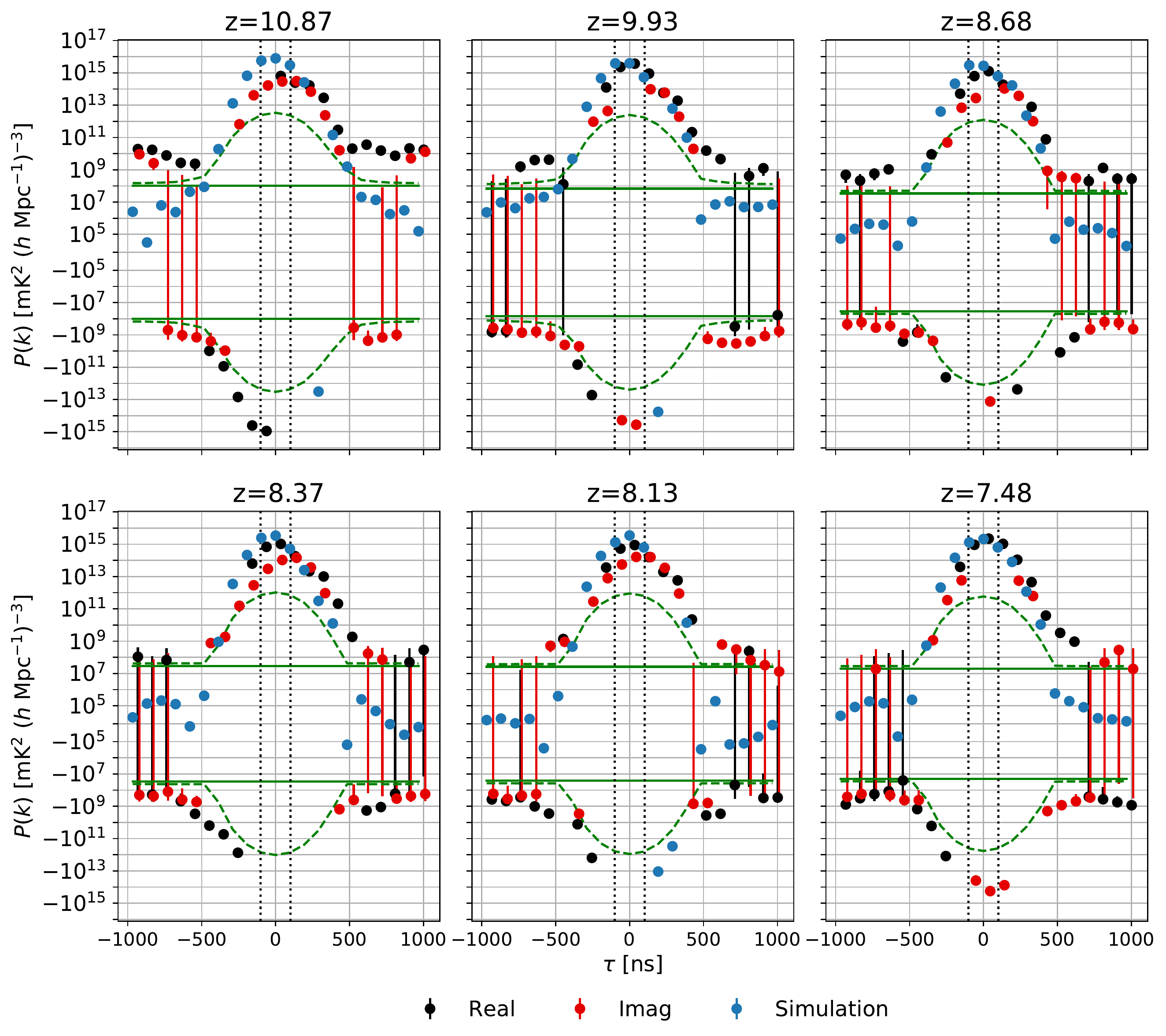}
\caption{Null-tests constructed by splitting the LST range ($[00^{h}30^{m}00^{s}, 08^{h}36^{m}00^{s}) $) in half (at $04^{h}30^{m}$), making two
power spectrum estimates and differencing the result. Real (black) and imaginary (red) are both shown, along with the null-test results
when applied to the simulated data (blue).
Such a difference would remove
isotropic cosmological signals leaving anything with dependence on sidereal time. Noise curves are as described in Figure~\ref{fig:pspec_vs_sim}. Statistically significant detections in the real part suggest power varying across the sky while significant imaginary power suggests a time dependence to phase calibration errors. The observed variations are consistent with simulation up to delays of 400\,ns.
The detections higher delay modes indicate large 
LST dependence which is inconsistent with cosmological power.}
\label{fig:lst_null_test}
\end{figure*}

\begin{figure*}[tp]
\centering
\includegraphics[width=\textwidth]{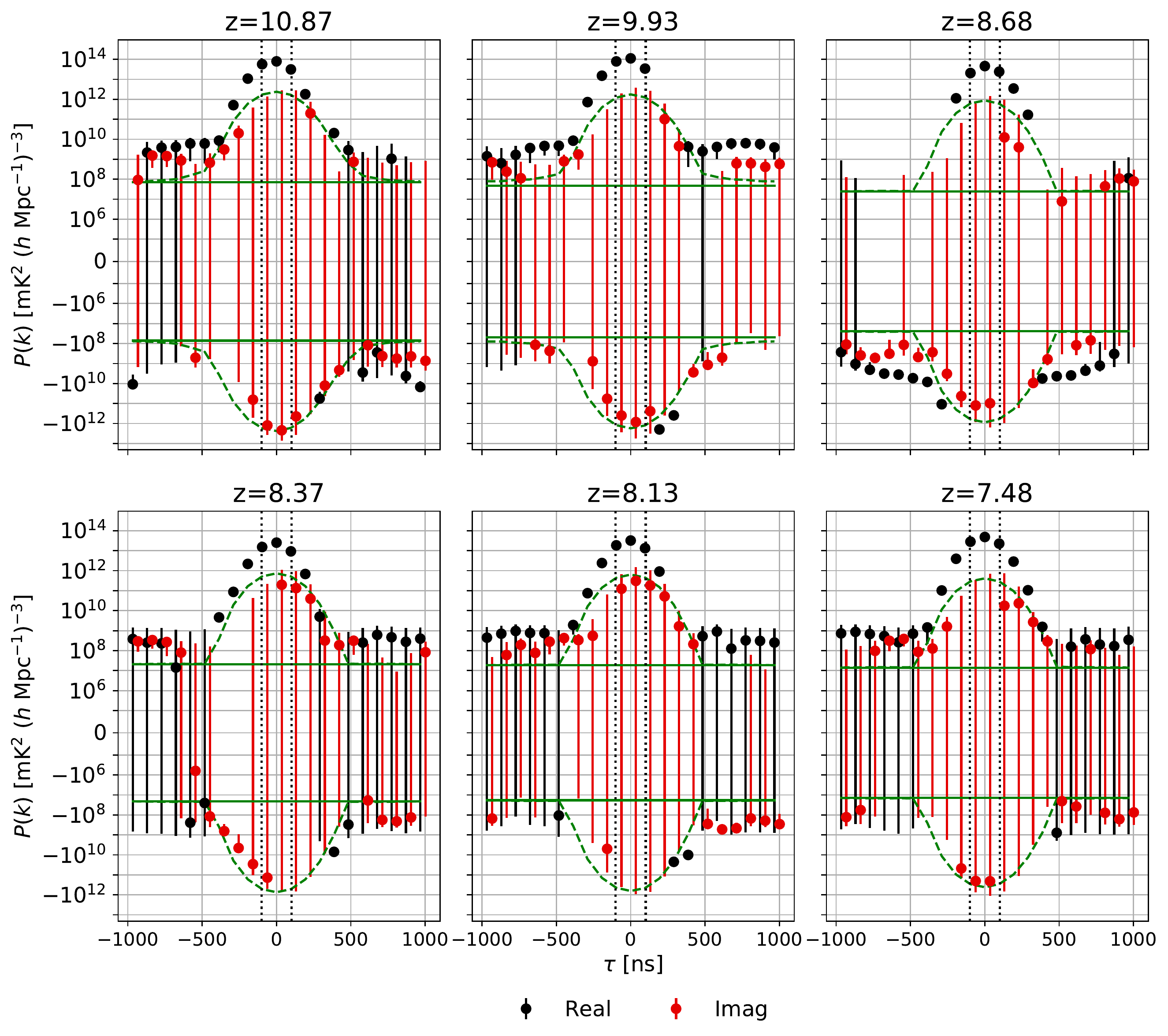}
\caption{In the LST binning process, data were split and binned into sets containing only even or odd numbered days; plotted here is the difference between the power spectra from these two sets. We use the same color scheme as Figure~\ref{fig:lst_null_test}. Where the largest difference in the LST null-test
(Figure \ref{fig:lst_null_test}) was on the order of 10\% of the measured value, here differences are less than 1\% at delays less than 400\,ns, and the imaginary points are nearly all consistent with predicted error bars. At delays larger than 400\,ns, statistically significant detections
in the three highest redshift bands are at comparable
levels to the power spectrum values in Figure~\ref{fig:pspec_vs_sim}.
This may be the result of contamination in only one set of the even or odd data (positive value for even, negative values for odd) which are mitigated during the cross-multiplication of these sets during power spectrum estimation. A variation in calibration as a function of JD can also cause the excess at delays less than $ 400 $~ns: days calibrated with the same solutions, but actually possessing some night-to-night gain variations, will result in some non-redundant signals between days.}
\label{fig:even_odd_null_test}
\end{figure*}

\subsubsection{Null-tests}\label{sec:nulls}
While the imaginary power suggests at least some presence of calibration error or non-redundancy, it does not fully explain the
origin of the excess power at delays greater than 400\,ns. Calibration errors, as long as they do not introduce
spectral structure, should not necessarily scatter power to high delays. Null-tests --- i.e. differences between
power spectra of different data selections --- can provide hints at the origin of these detections.

 For example, differencing the power spectra of two distinct stretches of sidereal time will remove
isotropic cosmological signals\footnote{Cosmological signals can only be removed through this method up to cosmic variance. However, since the thermal uncertainties dominate cosmic variance it is a decent approximation for this work.} but leave signals with strong dependence on sidereal time (like foregrounds).
Dividing the data set in half by LST into ranges $ [00^{h}30^{m}00^{s}, 04^{h}30^{m}00^{s}) $
and $ [04^{h}30^{m}00^{s}, 08^{h}36^{m}00^{s}) $ creates two sets of roughly equal sensitivity.
The resulting differenced power spectrum is shown in Figure~\ref{fig:lst_null_test}, along with a matching calculation for the foreground simulation.
The two are broadly consistent at
delays less than 400\,ns, i.e., they have the same the same sign and a
similar amplitude. Galactic synchrotron emission and bright point sources (like Fornax A
and Pictor A) are the most obvious contenders for strong variability.  We also see that the significant power seen in modes well beyond the horizon (for example, the strong positive offset at redshift 9.93 seen in the Figure~\ref{fig:pspec_vs_sim} power spectrum), is reflected in this null-test.

We also see that the imaginary component of the power spectrum null-test is comparable to the real component at most delay modes across all redshifts.
This suggests a sidereal time dependence of phase differences between baselines. In particular, note that the strong bias seen at redshift 9.93 is associated with a strong imaginary bias, implying  a phase rotation between baselines. Such an LST dependence of the imaginary component might be expected for non-redundancy (slightly different sky seen by nominally redundant baselines) or repeatable differences in calibration  which depends on the sky configuration (for example, one calibration solution when Fornax is transiting and a different one for when Pictor dominates.). This kind of variation in redundant calibration with sky flux density was shown in \citet{joseph_et_al2018}. 
Variations in calibrations from ionospheric fluctuations can also impact power spectrum estimation by introducing spectral structure and non-redundant information \citep{cotton_2004,intema_2009}.
This picture of non-redundancy strengthens the earlier hints  provided by Section~\ref{sec:redundancy}'s z-score analysis, which suggested that redundancy was particularly low around $120-130$\,MHz (redshifts 9 and 10).

A second easily constructed null-test is to difference
power spectra made from only the even and odd binned data sets.
Recall that these sets were constructed by separating even and odd
numbered days during the LST binning. A significant difference in this test would be suggestive of
a variation at the night-to-night level which departs significantly from the mean,
as these two sets are otherwise
expected to have identical sky signals
with different realizations of noise.

The resulting differenced power spectra for each redshift band
are show in Figure~\ref{fig:even_odd_null_test}.
Across all redshifts, there are points well beyond the
horizon which are inconsistent with both the analytic purely thermal variance and the foreground dependent uncertainty.  However there are two important differences
from the LST null-test. First, the overall amplitude of the difference power spectrum is much less. Within the horizon, the
difference amplitude is at most few $\times10^{13}$, or less than 0.1\% of the power spectrum.
Second, the imaginary power spectrum is consistent with
uncertainty across most modes. This is particularly notable within the horizon where even a small percent difference would drive a significant deviation. This
 suggests that  whatever causes the small but detectable difference between
 even and odd is not attributable to a phase difference between baselines.
 A variation in calibration as a function of JD can also cause the excess at delays less than $ 400 $~ns: days calibrated with the same solutions, but actually possessing some night-to-night gain variations, will result in some non-redundant signals between days.

The two highest redshift
bins again show the most significant differences at high delay;
the observed power values in this test are comparable to or even
exceed the power spectrum estimates shown in Figure~\ref{fig:pspec_vs_sim} and the imaginary leakage is $10$\% of that.
This result may provide  evidence of a signal
contaminating a single day that is averaged into the LST binned data set which is suppressed during the cross multiplication of days during power spectrum estimation.
Examples of such a systematic are improperly flagged RFI, a low amplitude signal not detected before cross-multiplication, or large transient gain isolated to a single night.

Another interesting feature can be seen in the redshift 8.68 bin in Figure~\ref{fig:even_odd_null_test}.
Here we see a consistent bias
which was not present in the mean power spectrum (Figure~\ref{fig:pspec_vs_sim}).  However, there is
a similarly shaped bias in the \emph{imaginary} part of the mean power spectrum.  A plausible hypothesis
is that, in this part of the spectrum, phase error between baselines is larger in one of the even/odd LST binned sets than the other.
However there is no clear significant difference in redundancy seen in the z-score/MAD analysis, so further evidence would be required to support this conclusion.

\subsubsection{Null-test Discussion}
Our two null-tests provide evidence that the foregrounds,
which vary significantly as a function of LST, are likely the cause of
some of the residual power detected at high delays during power spectrum estimation. There is also  some
evidence that suggests significant phase differences exist between
nominally redundant baselines, which introduce non-redundant signals
into the power spectrum estimates.

The presence of highly significant detections in the
even-odd null-test also suggests
there may be some net non-redundant signal between the
two LST binned data sets. These points are significant
compared to the propagated error bar ($ \sim10\sigma $ to $ \sim100\sigma $ inside the horizon) but
represent a small fraction of the total power observed ($ \leq 1\% $ of the power in Figure~\ref{fig:pspec_vs_sim}).
However the agreement of the
imaginary part of the power spectrum with the
foreground-dependent error bar suggests that each of the even-odd sets
has internally redundant baselines but the
data sets themselves are slightly different.

Both the null-tests discussed in this work, and the
presence of a significant fraction ($ \sim20\% $)
of power leaking from the real to imaginary component of
the power spectrum, indicate the presence of
non-redundant and non-isotropic signals. The latter
is not surprising since this analysis is performed
on data with no foreground subtraction and the sky
varies with LST as the galaxy and strong point sources
rise and set over an observation. In some places, particularly at low frequencies, this power
 couples to larger delays, presumably because of
 instrumental spectral structure. The even-odd null-test suggests that this
 spectral structure potentially varies in time while the imaginary component
 suggests that the spectral structure is not the same across nominally redundant baselines.

\subsection{Possible Future Directions}
\subsubsection{Jackknives in LST Binning}
An additional jackknife could be used to identify and possibly 
remove residual RFI and night to night variations identified in the even-odd null-test.
The variation is significant enough to be observable after differencing data averaged over the entire season.
If a specific night is the source of this result, it could potentially be further tracked down with additional jackknives  with  smaller sets of binned days or performing a null-test
by differencing data from the first and second half of the
observing season.
 This would provide information about the stability
of antennas and observations over the life of the
$ \PAPER $ experiment. Unfortunately, returning to the initial raw visibility data set is outside the scope of this analysis.
%Unfortunately data from pre-LST binning stages is not available during this analysis.

\subsubsection{Beam Null-test}
Non-redundancy happens when baselines, which in theory should see the same sky, in fact measure slightly different skies. Two obvious ways for this to happen are variations in antenna position and variation in beam pattern. In theory an element like PAPER should produce a symmetric beam, though this is not true in practice.
A simple test for non-redundancy due to beam differences
would be to test for deviations from symmetry by recording
observations with antennas rotated by 180\arcdeg. Differencing the 0\arcdeg and 180\arcdeg data sets would highlight abnormalities in the
beam response to the sky between antennas.
For an ideal, symmetric beam, all sky signal will
cancel and leave thermal noise fluctuations at all
times; however imperfections in beam response
will not cancel, resulting in a net signal in
the visibility data. Characterizing these net signals can help inform more precise beam models and place constraints on the level of beam-to-beam variation between different antennas.

\begin{figure*}[tp]
\centering
\includegraphics[width=\textwidth]{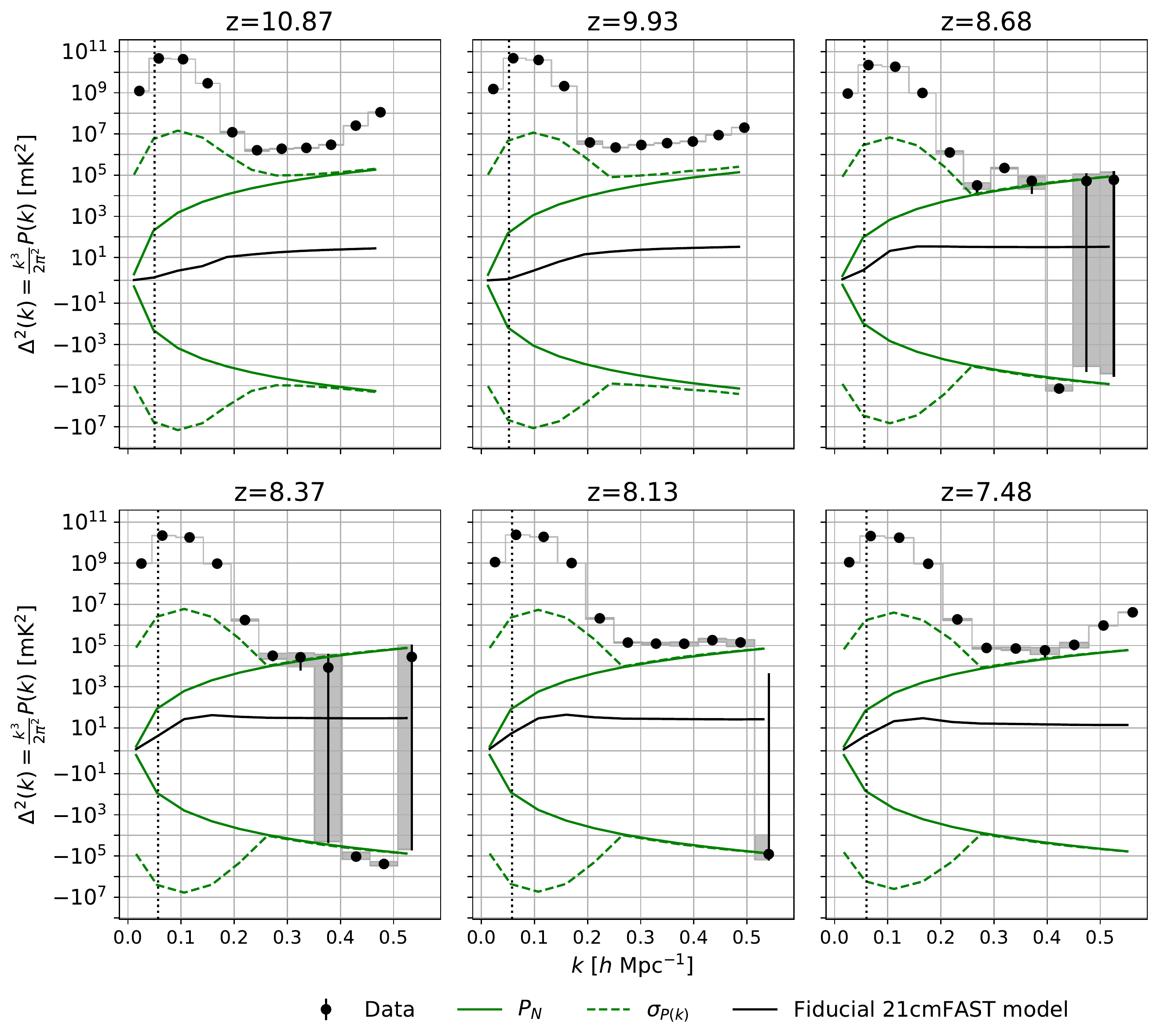}
\caption{The dimensionless power spectrum ($\Delta^{2}(k)= \frac{k^{3}}{2\pi^{2}}P(|k|)$) estimates and their uncertainties derived from the PAPER-64 observations.
All error bars represent $ 2\sigma $ uncertainties. Also plotted are the theoretical thermal noise limits from Equation~\ref{eqn:theory_noise} (solid green) and the foreground-dependent variance estimate from Equation~\ref{eqn:prop_error} (dashed green line).
These are the same lines as in Figure~\ref{fig:pspec_vs_sim} with the addition of the
black solid line representing fiducial \code{21cmFAST} models of reionization for comparison.
The horizon line (vertical dotted black) has been transformed from the maximum
signal delay between antennas to the cosmological co-moving size scales using Equations 12 and 13  of \citet{liu_et_al2014a}. }
\label{fig:pspec_delta2}
\end{figure*}

\section{21cm Upper Limits}\label{sec:upperlims}
We use the PAPER data to place upper limits on the 21cm signal strength using the
dimensionless power spectrum: $\Delta^{2}(k)= \frac{|k|^{3}}{2\pi^{2}}P(|k|)$. To convert from
interferometric delay to cosmological co-moving wavenumber, we assume Planck-15  cosmology.
 These power spectra are shown in Figure~\ref{fig:pspec_delta2}.

As a summary and comparison of progress across the field, we also report, from each published power spectrum, the lowest upper limits in each redshift band, shown in Figure \ref{fig:eor_summary}.
This minimum is taken across the $k$ ranges reported by each 
experiment to be free of possible signal loss or other extraneous
factors (for example, early PAPER results reported values inside the filtered wedge but indicated they were not to be used).

To encapsulate the results of this work, the most sensitive limit is reported from
the range $0.3 <  k < 0.6\ \hMpci$, where both null-tests
pass for most $k$-modes in each redshift bin.
These limits on the 21cm power spectrum
from reionization are \upperlims. 
Table~\ref{tab:summary} also provides a summary of this data.

 These upper limits represent a significant
increase compared to prior limits published by the PAPER instrument (a factor of $\sim10$ in mK). They also exceed the expected
amplitude of a fiducial \code{21CMFAST}\footnote{\url{github.com/andreimesinger/21cmFAST}}
 model by a factor of $\sim100$ in mK \citep{mesinger_et_al2011}. These limit supersede all
 previous PAPER results for reasons described in \citetalias{cheng_et_al2018}.

\begin{deluxetable}{cccc}[tp]
\tablewidth{.5\textwidth}
\centering
\tablecolumns{4}
\tablecaption{The minimum volumed weighted power spectrum estimates $ \Delta^{2}(|k|) $~[mK]$ {}^{2} $ from this analysis computed over the range $ .3 < |k| < .6 $}
\tablehead{
    \colhead{Redshift}  &
    \colhead{$ |k| $} &
    \colhead{$ \Delta^{2}(|k|) $} &
    \colhead{$ \delta  \Delta^{2}(|k|)$\tablenotemark{a}}\\
    \colhead{} &
    \colhead{[$ h $/Mpc]} &
    \colhead{[mK]${}^{2}$} &
    \colhead{[mK]${}^{2}$}
}
\startdata
7.49 & 0.39 & 5.6$\times 10^{4}$ & 3.5$\times 10^{4}$ \tabularnewline
8.13 & 0.32 & 1.2$\times 10^{5}$ & 2.0$\times 10^{4}$ \tabularnewline
8.37 & 0.37 & 1.0$\times 10^{4}$ & 3.2$\times 10^{4}$ \tabularnewline
8.68 & 0.36 & 3.8$\times 10^{4}$ & 4.1$\times 10^{4}$ \tabularnewline
9.93 & 0.34 & 3.5$\times 10^{6}$ & 1.9$\times 10^{5}$ \tabularnewline
10.88 & 0.33 & 2.1$\times 10^{6}$ & 1.5$\times 10^{5}$
\enddata
\tablenotetext{a}{All uncertainties are 2$ \sigma $}
\label{tab:summary}
\end{deluxetable}

\section{Conclusion}\label{sec:conclusion}
 We have re-analyzed the PAPER-64 data first presented in \citetalias{ali_et_al2015},
 and presented 21cm
power spectra and uncertainties in 5 independent redshift bins. These estimates
are made using an independently developed pipeline
which skips foreground subtraction and simplifies time averaging. Simulations
of noise and foregrounds are used to build a basic picture of
internal consistency. The resulting power spectra reach the noise limit
across much of the spectrum but above redshift $9$ (below $130$\,MHz), they demonstrate
a statistically significant excess of power.  Null-tests support a picture where
these power spectrum detections are caused by foregrounds modulated by
spectrally dependent deviations from redundancy or calibration error.
In particular, the
z-scores and imaginary power tests suggest that residuals
could be the result of some net non-redundant signal between 
baselines in a nominally redundant group.

Future analyses of highly redundant sky
measurements will require strict comparisons
between nominally redundant samples before
cross-multiplication to ensure effects like
these can be mitigated. Also further jackknives
and comparisons of data should be done before or as part of LST binning
to detect likely contributions to excess.
They will also require more precise antenna placement
to ensure baselines designed to be redundant do not
introduce signal in the imaginary component of the power spectrum.

These results represent the
most robust results from the PAPER experiment and supersede all previous PAPER
power spectrum limits. This includes results both from PAPER-32 \citep{parsons_et_al2014,jacobs_et_al2014,moore_et_al2017},
which used a different covariance estimation technique
 but have not been subjected to a rigorous re-analysis
\`{a} la \citetalias{cheng_et_al2018}, and previous
PAPER-64 results \citep{ali_et_al2015,ali_et_al2015_erratum}.
Any constraints on the spin temperature of hydrogen made by
\citet{pober_et_al2015} and \citet{greig_et_al2015} based on the
previously published upper limits should also be disregarded.
Though these measurements do not place significant constraints on the IGM
temperature, the analysis presented in these two
papers remains relevant to any future limits on the 21cm power spectrum 
at levels similar to the original results of \citetalias{ali_et_al2015}.

\begin{figure*}[tp]
\centering
\includegraphics[width=.95\textwidth]{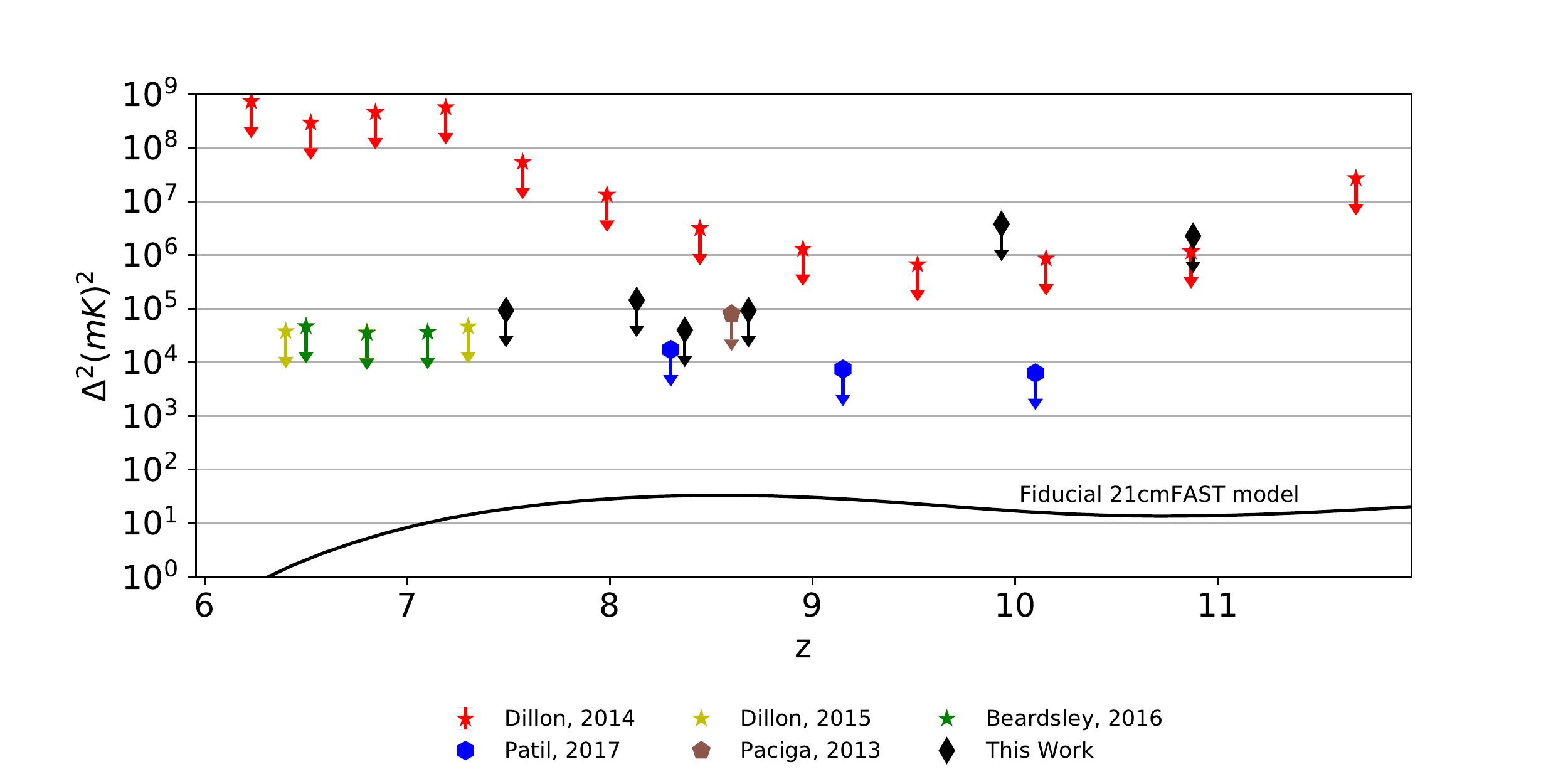}
\caption{A comparison of the lowest limits achieved by various instruments in the $k$-ranges reported by each instrument.
The results reported from this paper are taken in the range 
$0.3 \leq k \leq 0.6\ \hMpci$. Data is taken
from the MWA (stars; \cite{dillon_et_al2013b,dillon_et_al2015,beardsley_et_al2016}),
the GMRT (pentagon; \cite{paciga_et_al2013}),
LOFAR (hexagons; \cite{patil_et_al2017}),
and PAPER (diamonds; this work).
We include the $z=8.37$ redshift bin analyzed both here and in \citetalias{cheng_et_al2018}, although
it is worth noting this redshift bin is not entirely independent
from the $z=8.13$ and $8.68$ bins, as can be inferred from the
overlapping window functions from Figure~\ref{fig:freq_select}.
For reasons described in \protect{\citetalias{cheng_et_al2018}},
these PAPER results should supersede all previous PAPER limits.
\label{fig:eor_summary}}
\end{figure*}

The current best limits from 21cm power spectrum experiments 
 are shown in Figure~\ref{fig:eor_summary}.
To date, all power spectrum estimates have been reported as
upper limits. However, to discern and characterize the physics
of reionization, high significance detections of the 21cm power
spectrum are necessary.
Next generation radio telescopes,
like the fully realized 350 element configuration of HERA
\citep{pober_et_al2014,DeBoer_et_al2016,liu_parsons2016}
and the
future Square Kilometre Array (SKA; \citet{mellema:2013}), are predicted to
be able to make these detections and put stringent constraints
on reionization.

\section{Acknowledgments}
We would like to thank Adam Beardsley, Judd Bowman, Bryna Hazelton, and Miguel Morales for their insightful discussions. We would also like to
 thank Ruby Byrne and Patti Carroll for supplying extended source models of Fornax A and Pictor A.

This work made use of the following python software packages: pyuvdata \citep{pyuvdata}, astropy \citep{astropy_2013},
scipy \citep{scipy}, and numpy \citep{numpy}. Some of the results in this paper have been derived using the HEALPix \citep{gorski_et_al2005} package.

MJK is supported by the NSF under project number AST-1613973
and would also like to acknowledge the support of Arizona State University.
CC would like to acknowledge the UC Berkeley
Chancellor’s Fellowship, National Science Foundation
Graduate Research Fellowship (Division of Graduate Education award 1106400).
PAPER and HERA are supported by
grants from the National Science Foundation (awards
1440343, and 1636646). ARP, DCJ, and JEA would
also like to acknowledge NSF support (awards 1352519,
1401708, and 1455151, respectively). SAK is supported
by the University of Pennsylvania School of Arts and
Sciences Dissertation Completion Fellowship.
AL acknowledges support from a Natural Sciences and Engineering Research Council of Canada (NSERC) Discovery Grant and the Canadian Institute for Advanced Research (CIFAR) Azrieli Global Scholars program.
GB acknowledges the Rhodes University research office, funding from the INAF PRIN-SKA
2017 project 1.05.01.88.04 (FORECaST), the support from the Ministero degli Affari
Esteri della Cooperazione Internazionale - Direzione Generale per la Promozione del
Sistema Paese Progetto di Grande Rilevanza ZA18GR02 and the National Research
Foundation of South Africa (Grant Number 113121) as part of the ISARP
RADIOSKY2020 Joint Research Scheme.
This work is based on the research supported in part by the National Research Foundation of South Africa (grant No. 103424).
We would also like to thank SKA-SA for site infrastructure and observing support.

\software{simpleDS 
    (\url{github.com/RadioAstronomySoftwareGroup/simpleDS}),
    pyuvdata \citep{pyuvdata},
    scipy \citep{scipy},
    numpy \citep{numpy},
    astropy \citep{astropy_2013},
    PRISim \citep{prisim}
}

\bibliographystyle{aasjournal}
\bibliography{biblio}

\appendix
\section{Foreground Dependent Variance}\label{appendix:var}
To find the variance of $ P(k) $, being by assuming each
visibility $ \tilde{V}_{i}(\tau,u,v,w)  = s + n_{i}$ is
the sum of the true sky signal, $ s $ , and a noise
component $ n_{i} $. 

For convenience define the cosmological conversion factor
\begin{align}
\Phi = \left(\frac{\lambda^{2}}{2k_{B}} \right)^{2} \frac{X^{2}Y}{\Omega_{pp}B_{pp}}
\end{align}
Also for simplicity in this analysis, we 
ignore cosmic variance in the signal term. This results 
in the signal term being not a random variable but related
to the power spectrum of the sky by 
$ s^{2} = P_{s}(k) / \Phi $,  where $ P_{s}(k) $ is the true power spectrum of the sky signal for a delay transformed visibility.
Let the noise term be drawn from the complex distribution $ n_{i} \sim \mathcal{CN}(0, \sqrt{P_{N}(k)/\Phi}) $ where $ n_{i} $ is independent for each baselines\footnote{This assumption does ignore the correlations induced between visilibities that
share a common antenna and thus have correlated noise.}.

Then we can propagate the variance in $ P(k) $ as

\begin{align}
	\sigma^{2}_{P(k)}=Var\left(P(k)\right) &= Var\left(
	\Phi
	\left\langle \tilde{V}^{\star}_{i}(\tau,t)\tilde{V}_{j}(\tau,t) \right\rangle_{i\neq j,LST}\right)\label{eqn:approximation_expect}\\
	& = \left(\Phi \right)^{2}
	\left\langle Var\left( \tilde{V}^{\star}_{i}(\tau,t)\tilde{V}_{j}(\tau,t)\right) \right\rangle_{i\neq j,LST}\\
	& =\left(\Phi \right)^{2}
	\left\langle Var\left( (s + n_{i})^{\star}(s + n_{j}\right) \right\rangle_{i\neq j,LST}\\
	& = \left(\Phi \right)^{2}
	\left\langle Var\left(s^{2} + sn^{\star}_{i} + sn_{j} + n_{i}^{\star}n_{j}\right) \right\rangle_{i\neq j,LST}\\
	& = \left(\Phi \right)^{2} \left\langle  s^{2}Var(n_{i}^{\star})  + s^{2}Var(n_{j}) + Var(n_{i}^{\star}n_{j })\right\rangle_{i\neq j,LST}\\
	&= \left(\Phi \right)^{2}
	\left\langle \left( 2s^{2}Var(n_{j}) + Var\left(n_{i}^{\star}n_{j}\right)\right) \right\rangle_{i\neq j,LST}\\ \label{eqn:app_expect}
	&=  \left(\Phi \right)^{2}	\left\langle \left( 2s^{2}Var(n_{i}) + E\left[n_{i}n_{i}^{\star}{}n_{j}n_{j}^{\star}\right] - \Big|E\left[n_{i}^{\star}n_{j}\right]\Big|^{2} \right) \right\rangle_{i\neq j,LST}\\
	&=  \left(\Phi \right)^{2}	\left\langle  2s^{2}Var(n_{i}) + E[|n_{i}|^{2}] E[|n_{j}|^{2}]\right\rangle_{i\neq j,LST}\\
	&=  \left(\Phi\right)^{2}	\left\langle  2s^{2}Var(n_{i}) + Var(n_{i})^{2}\right\rangle_{i\neq j,LST}\\
	&=  \left(\Phi \right)^{2}	\left\langle  2\frac{P_{s}(k)P_{N}(k)}{\Phi^{2}} + \frac{P_{N}(k)^{2}}{\Phi^{2}}\right\rangle_{i\neq j,LST}\\
	&= \left\langle  2P_{s}(k)P_{N}(k) + P_{N}(k)^{2}\right\rangle_{i\neq j,LST}
\end{align}
Where we have assumed each $ n_{i} $ are independent
random variables as mentioned above and all constants
of proportionality have been used to transform the power spectra from functions of delay, $\tau $, to cosmological wavenumber, $ k $. 
This derivation assumes noise is independent on the baseline level.
It also assumes the power spectrum and noise are independent
in time. In general, these assumptions may not be true, and would contribute to additional covariance terms in the expansion of the ensemble average in Equation~\ref{eqn:approximation_expect}.

At high delay modes, foreground signals are predicted to 
have little power (e.g. $P_{s}(k)\to 0 $)
 and the variance reduces to the thermal
variance $ P_{N} $. Conversely inside the horizon, and at
delay modes just outside the horizon, this variance will
be dominated by the term dependent on the power spectrum of the true sky $ P_{s}(k) $.

\listofchanges
\end{document}